\documentclass[10pt]{reportMaster}

\usepackage{graphicx}
\usepackage{natbib}
\usepackage{float}
\usepackage{caption}
\usepackage{subcaption}
\usepackage{xcolor,amsmath}
\usepackage{amsmath,amssymb}
\usepackage{makecell}
\usepackage[toc,page]{appendix}

\usepackage[linesnumbered,ruled,vlined]{algorithm2e}
\DontPrintSemicolon

\SetKwComment{Comment}{\color{green!50!black}// }{}

\SetKwProg{Function}{function}{}{}

\title{Exploring constraints on CycleGAN-based CBCT enhancement for adaptive
radiotherapy}
\author{B. S. \ Pai}
\id{DKE-21-38}
\nameMastersProgramme{Artificial Intelligence}
\committee{Dr. E. Hortal Quesada\\ Dr. S. Asteriadis \\ Dr. I. Zhovannik}
\date{\today}

\begin{document}
\maketitle

\begin{abstract}
Research exploring CycleGAN-based synthetic image generation has recently accelerated in the medical community, as it is able to leverage unpaired datasets effectively. However, clinical acceptance of these synthetic images pose a significant challenge as they are subject to strict evaluation protocols. A commonly established drawback of the CycleGAN, the introduction of artifacts in generated images is unforgivable in the case of medical images. In an attempt to alleviate this drawback, we explore different constraints of the CycleGAN along with investigation of adaptive control of these constraints. The benefits of imposing additional constraints on the CycleGAN, in the form of structure retaining losses is also explored. A generalized frequency loss inspired by \cite{jiang2020focal} that preserves content in the frequency domain between source and target is investigated and compared with existing losses such as the MIND loss~\citep{yang2018unpaired}. Synthetic images generated from our methods are quantitatively and qualitatively investigated and outperform the baseline CycleGAN and other approaches. Furthermore, no observable artifacts or loss in image quality is found, which is critical for acceptance of these synthetic images. The synthetic medical images thus generated are also evaluated using domain-specific evaluation and using segmentation as a downstream task, in order to clearly highlight their applicability to clinical workflows.

\end{abstract}

\tableofcontents
\chapter{Introduction}

\section{Medical Imaging and Adaptive Radiotherapy}
Medical imaging plays a key role in providing a simulation of the human body through imaging techniques for aiding multiple medical procedures such as diagnosis, prognosis and treatment. Many different imaging techniques form a part of medical imaging aiding radiologists/clinicians to gain additional information that may not be available solely with the naked eye. This additional information can provide key insight to a radiologist for decision making \citep{Doi_2006}. More specifically for oncology, medical imaging is essential in order to detect cancer earlier and provide improved outcomes to the patient. 

Several medical imaging techniques (modalities) are used for diagnosis of cancer such as Computed Tomography (CT), Positron Emission Tomograpy (PET), Magnetic Resonance Imaging (MRI), etc. Each of these modalities have unique properties and provide specific information due to the physics of their acquisition. These can then be leveraged accordingly to support different stages of decision-making. Medical Imaging also plays a key role in treatment of cancer where different modalities can be used to design a course of action. Surgery, chemotherapy and radiotherapy can all benefit by being driven through imaging~\citep{Frangioni2008}.  

\subsubsection{Radiation Therapy}
Radiotherapy is a method of treatment of cancer where doses of radiation are targeted at the tumour in order to stop its growth and kill tumour cells. While delivering dose to the tumour, healthy organs need to be spared in order to avoid causing damage to them which makes the design of radiotherapy treatment a challenging task. Radiotherapy can be either 
\begin{enumerate}
    \item External Beam Radiotherapy (EBRT):
    A machine placed externally delivers beams of radiation targeted at the tumour. 
    \item Internal Radiotherapy (IRT): A radioactive source is placed inside the body and delivers radiation internally. 
\end{enumerate}
EBRT consists of a large set of techniques that can rely on different types of beams such as proton or photon beams. 3D conformal radiotherapy relies on medical imaging to deliver radiation to the tumour by targeting beams at it from multiple directions. Intensity Modulated Radiotherapy (IMRT) is a more advanced technique where said beams are shaped more precisely to conform to the shape of a tumour. In addition to precisely shaping beams, radiation therapy can also be improved by leveraging medical images not just while planning but also during treatment delivery in a method called Image Guided Radiotherapy (IGRT). In an IGRT workflow, images are used during treatment delivery to position the patient accurately or to change the radiation dose. IGRT can help mitigate errors induced during initial imaging during planning and account for treating sites where motion might be common such as the lung \citep{LING2006119}. More recently, the capturing of these \textit{guidance} images have allowed adapting radiotherapy plans during the process of treatment. 

\subsubsection{Adaptive Radiotherapy}
Adaptive Radiotherapy (ART) is a set of emerging data-driven techniques that aims to administer more accurate treatment plans that account for changes observed during the process of treatment planning and delivery. Multiple factors can lead to changes being observed during the treatment process such as,
\begin{itemize}
    \item Geometrical uncertainties: \\
    These are uncertainties are caused by both systemic and random errors. Systemic errors such as delineation differences, setup errors and organ motion can cause reduction in accuracies of the RT process. 
    \item Anatomical changes: \\
    Tumour regression along with changes in the size/shape of the different organs over the course of treatment can lead to differences in ideal dose versus the one estimated during initial planning.
    \item Biological response: \\
    Changes in the biology of the tumour can also occur during treatment which may contribute to changes in toxicities for surrounding organs at risk. 
\end{itemize}
\cite{SONKE201094} discuss the above change inducing factors in detail. They propose that adaptive methods in RT can help account for these changes to a certain extent by repetitive imaging and analysis.

\subsubsection{ART Process Workflow}
\begin{figure}[H]
    \centering
    \includegraphics[width=0.9\textwidth]{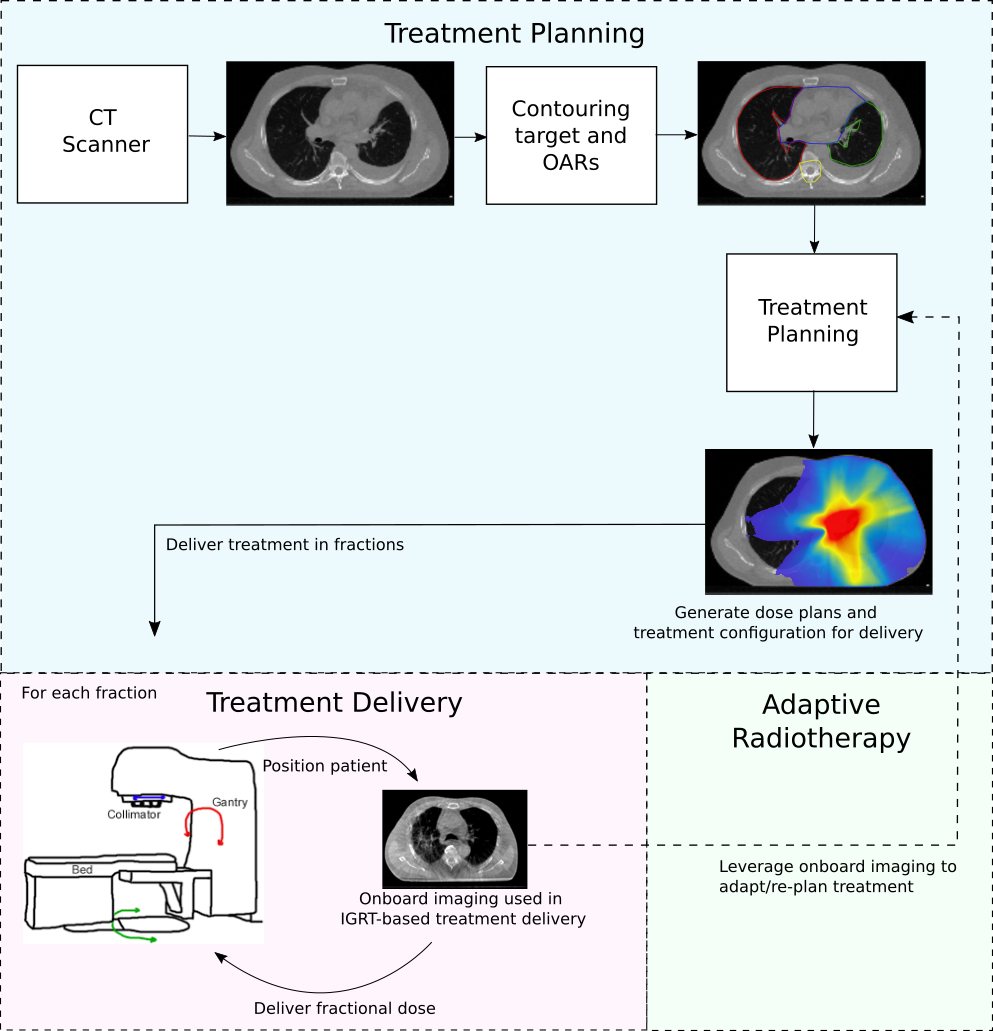}
    \caption{ART Process Workflow showing different processes of treatment planning, treatment delivery and adaptive radiotherapy.}
    \label{fig:intro:adaptive_rt}
\end{figure}

Figure \ref{fig:intro:adaptive_rt} shows a high level overview of the treatment planning and delivery process with integration of adaptive radiotherapy. The treatment planning process begins by the acquisition of a CT scan, followed by contouring of the target (tumour, lymph nodes and other malignant locations) and organs-at-risk (OAR) by radiation therapists. The CT scan along with contours are used to generate treatment plans and dose distributions that can then be delivered using a linear accelerator. The treatment plan is generally designed, such that, it is administered across different fractions or sittings. At each fraction, a fractional dose is delivered to the patient. When IGRT-based methods are leveraged, on-board imaging such as CBCT or MRI is used to position the patient accurately before delivering the prescribed dose. These modalities are preferred as they provide either minimal or no radiation to the patient. This process is repeated across multiple fractions. Adaptive radiotherapy, as shown in the green block in Figure \ref{fig:intro:adaptive_rt}, suggests using on-board imaging data to adapt/re-plan treatment as required. This allows improving treatment outcomes by providing minimal overdosing to healthy tissues while concentrating maximum dose to the tumour. However, on-boarding imaging 

\subsection{Significance of CT and CBCT imaging}\label{sec:intro:cbctvsct}
Computed Tomography (CT) was the first method that provided non-invasive imaging of the human body without being biased by overlapping structures owing to its 3D nature \citep{Buzug2011}. CT imaging uses X-ray radiation aimed at the body in a fan-like shape falling on a line detector to generate 3D cross-sectional images. The intensities within these cross-sectional images are dependent on the density of bone, tissues and organs within the body and therefore allow the observing the make-up of the body when stacked. These densities are represented as Hounsfield Units (HU) in a CT scan. In addition to providing beneficial visual information,  CT scans can be used for planning treatment as the modelling of densities allow gauging how the same objects might respond to radiation.
\par
\begin{figure}[H]
    \centering
    \includegraphics[width=\textwidth]{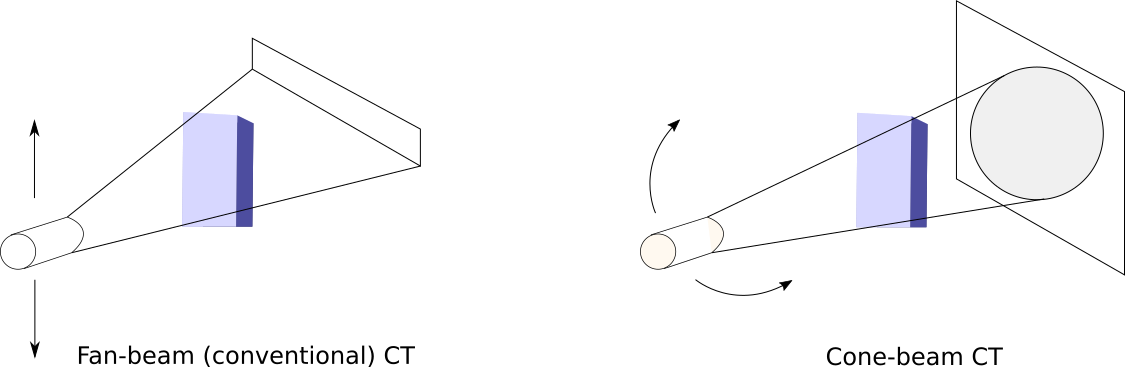}
    \caption{Acquisition physics of fan-beam and cone-beam CT. In fan beam CT, the source emits a fan-shaped X-ray beam on the object which is then collected by a line-detector. The source moves along the vertical axes capturing cross sectional imagery. In cone-beam CT, the source emits a cone-shaped beam that falls on a flat panel detector. The source revolves around the object and captures projections at different angles.}
    \label{fig:intro:cbct_ct_physics}
\end{figure}
\begin{figure}[H]
     \centering
      \includegraphics[width=\textwidth]{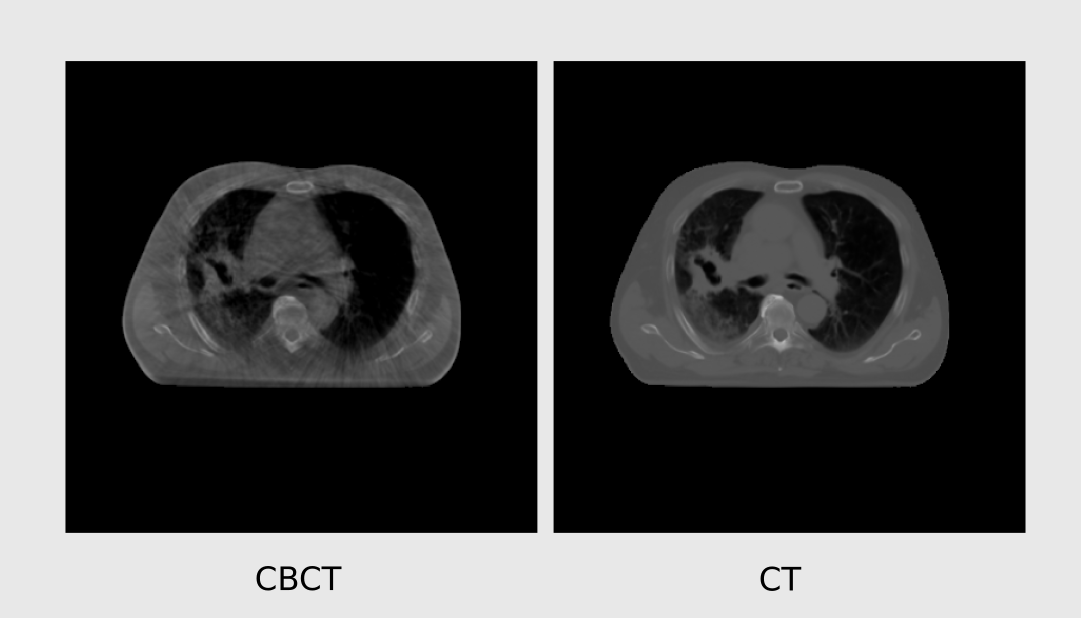}
    \caption{Mid-axial views of a patient scanned using cone-beam CT (CBCT) and fan-beam CT (CT). A rough glance at these images shows that the cone-beam image is much noisier with streaks dominant across the image. The Hounsfield Units (HU) of the cone-beam CT scan are not standardized either. Water, which would have a standard HU value of 0 in a fan-beam CT might not present the same value in a cone-beam CT.}
    \label{fig:intro:cbct_ct_examples}     
\end{figure}

Cone-beam Computed Tomography (CBCT) imaging uses a cone-shaped X-ray beam with a flat panel detector. Unlike conventional CT, CBCT generates 3D images by projecting cone-beams at different angles and reconstructing a 3D image using projection geometry. This makes the image more susceptible to scattering and noise when compared to a CT. However, due to smaller footprint and lower dose exposure of CBCT scanning systems, they have been integrated quite successfully as onboard imaging on linear accelerators \citep{Sterzing2011}. Both CT and CBCT images are crucial to the process of radiation therapy. CT allows generating treatment plans via the acquistion of a planning CT while CBCT allows efficient delivery by more accurate positioning and correction through IGRT. CBCT can also play a key role in adaptive radiotherapy providing better outcomes for the patient. Figure \ref{fig:intro:cbct_ct_physics} shows the acquisition physics of CBCT and CT scanners. A sample of resulting scans from these scanners are show in Figure \ref{fig:intro:cbct_ct_examples}.

\section{Image quality improvement for adaptive radiotherapy}\label{sec:intro:improvement}
CBCT images are are acquired with about an order of magnitude lesser radiation than conventional fan-beam CTs \citep{Schulze2011}. This renders them more susceptible to sources of noise that already affect CT imaging \citep{Lechuga2016}. In addition, physics of CBCT acquisition introduce sources of noise that fan-beam CTs generally do not face such as increased scatter due to usage of flat panel detectors. Combining both these factors, CBCT images often present with lower quality. \cite{Schulze2011} present a comprehensive overview of the different artefacts that affect the quality of a CBCT scan. Along with affecting the visual quality, these artefacts can also contribute to the dosimetric accuracy of using a CBCT scan. 
\par
\cite{Posiewnik2019} discuss CBCT image quality and highlight two groups of issues that can prove adaptive radiotherapy with CBCT challenging. The first group consists of image quality degradation that manifests as reduction of contrast, obscuring structures and impairment of regions of interest. This is mainly due to the various artefacts outlined by \cite{Schulze2011}. The second group of issues pertain to parameters that determine the CBCT acquisition. Standardized guidelines and QA procedures can lead to acceptable CBCT image quality whereas deviation from these can cause adverse effects. 
\par
Dose calculation capability or dosimetric quality of CBCT imaging is integral in determining its value in ART. Dosimetric quality is defined as the ability of the image to produce accurate dose distributions when used for treatment planning. Therefore, dosimetric quality similar to fan-beam CT imaging would be ideal with increasing deviations affecting the efficacy of ART procedures.  \cite{Schulze2011} highlight multiple studies that show that dose calculated on CBCT vs CT present unacceptable deviations when HU calibration is done through look-up tables. 
\par 
Considerable research has been presented on the benefit of processing CBCT images to improve their quality and reduce artefacts with traditional methods such as scatter correction \citep{Jin2010}, density overrides \citep{Dunlop2015}, CT number calibration \citep{Dunlop2015}, deformable image registration \citep{Landry2015} and certain model-based methods \citep{Zhao2016}. With the democratization of machine learning and deep learning, multiple studies have also presented ML/DL-based methods to post-process CBCT images. These methods are often much quicker and less cumbersome in contrast to traditional methods and have found wide acceptance in the medical imaging research community. However, in-terms of clinical implementation, the vulnerability of ML/DL-based methods in dealing with out-of-distribution data has been a limiting factor. In the section below, a short review of some deep learning approaches along with their benefits and limitations is discussed. 

\subsection{Deep Learning approaches}
Deep learning-based approaches for CBCT improvement have mostly revolved around encoder-decoder networks or generative-adversarial networks where voxel regression is performed with the original CBCT image as input. The vast majority of such approaches can be categorized into either paired or unpaired. 

\subsubsection{Paired approaches}
Paired approaches require input images with ground truth that denotes what the ideal output would be. This is generally a highly matching CT image as CBCT improvement is synonymous with translation to a CT like quality. A CBCT image along with its highly matching CT can be collected in various ways - planning CT and first positioning CBCT might provide required correspondences. CTs can also be taken during the treatment delivery process, called rescanned CTs, and generally correspond well with CBCTs taken for positioning around that time. The CBCT and its matching CT image need to be co-registered to align anatomies accounting for differences in scan parameters and movement. Upon obtaining images with GT pairs, these can then be used in a fully-supervised fashion. 
\par
Paired approaches using encoder-decoder networks, more specifically using UNets \citep{Ronnerberger2015} have been presented by multiple studies~\citep{Kida2018, Landry2019, Yuan2020, Thummerer2020}. \cite{Kida2018} train their model with 20 patients in a 2D fashion using a 39-layer UNet architecture. They show that their methods improve the image and dosimetric quality through better quantitative scores for metrics such as Structural Similarity Index \citep{wang2004image}, Power-to-Signal Noise Ratio, ROI Mean values and Spatial Non-Uniformity. \cite{Landry2019} compare three UNets trained with different input representations --- in projection and image spaces --- and aim to identify the best approach for dosimetric accuracy. They use L2 norm between the target and predicted images in contrast to MAE \citep{Kida2018}. They show promising results in both projection and image spaces concluding that both are feasible methods for CBCT correction. \cite{Yuan2020} present an approach using three cross-sectional slices as 3-channel inputs to a 2 dimensional UNet. The authors extensively validate their approaches through group-based cross-validation and show large improvements in image quality across all their test studies. \cite{Thummerer2020} train each 2D network considering different planes --- axial, sagittal and coronal --- of the 3D CBCT-CT scan pair followed by aggregation during test-time. Although they show the efficacy of their methods compared to traditional approaches, they do not present comparisons of single plane vs their multi-planar approach. \par
Image-to-image translation community placed a lot of importance on designing handcrafted losses between target and predicted image spaces in order to integrate domain knowledge and preserve relevant features in the predicted image. With the introduction of Pix2pix GAN \citep{Isola2016}, a discriminator was designed to substitute complex loss design and preserve high frequency information. CBCT to CT translation was also benefit by addition of a discriminator. \cite{Zhang2020} compare multiple deep learning approaches from UNet encoder-decoders to CycleGANs and show that their pix2pix GAN outperforms other methods which shows a clear benefit of the pix2pix framework. Do note that their CycleGAN was trained using paired data and does not leverage unpaired data. Another interesting work by \cite{Dahiya2021} show how physics based data-augmentation can be used to create paired data which can then be leveraged in a pix2pix framework. In addition to improved CBCT, they also generate orgam segmentations for the CBCT image. 

Analysis of the above studies shows that paired approaches are feasible candidates for image quality improvement. However, the pre-processing needed to obtain paired data might pose a hindrance in utilizing all available data efficiently. During treatment, multiple CBCTs are acquired but, generally, only a single pair can be formed when matching with a planning CT. Usage of rescanned CTs need additional scans to be taken which might be inconvenient for the patient due to increased cost, radiation and time. Moreover, the process of pairing the data might also introduce biases such as dependence on method/quality of registration chosen. These impeding factors along with the emergence of unpaired approaches such as CycleGAN has lead the research community to lean toward unpaired approaches. 

\subsubsection{Unpaired approaches}
These approaches are designed to handle data where matching characteristics may be hard to obtain. For example, in the task of converting horses to zebras, it is almost impossible to have a known ground truth of what a horse might look like if it were a zebra. We can consider a set of horse images to form the source data space and a set of zebra images to form the target data space. This is where unpaired approaches enter and leverage the properties of these data spaces and allow translation between them without enforcing strict matches. 
\par
The CycleGAN framework introduced by \citep{Zhu2017} is one of the most consistently used unpaired approaches for image-to-image translation. \cite{Kurz2019} present an approach using a 2D CycleGAN where co-registered slices are used as inputs. Their methods show very good correspondence with an existing CBCT correction method, both image and dosimetry-wise, while being much faster. \cite{Maspero2020} use limited field-of-view CBCTs and rescanned CTs in a purely unpaired fashion across three different anatomical sites. A single network trained on all sites along with individually trained networks are explored in their work. Both approaches perform similarly and show a large improvement in image similarity metrics when evaluated with a set of ground truth pairs. They additionally show that the improved CBCT is of sufficient dosimetric quality through dose differences and gamma analysis. \cite{Liu2020} demonstrate the use of attention gates in CycleGANs and show that their introduction improves smoothness and reduces artifacts when compared with a UNet and a vanilla CycleGAN. Various other studies by \cite{Harms2019, Eckl2020, Kida2020} show the benefit of CycleGAN approaches in CBCT improvement for both visual and dosimetric tasks.

\section{Cycle-consistent Generative Adversarial Networks}
In this section, an overview of Generative Adversarial Networks (GAN) is provided followed by an in-depth description of a variant of GAN for image-to-image translation in unpaired settings - the Cycle-consistent Generative Adversarial Network or CycleGAN. Constraints that are set in a CycleGAN are discussed and studies that discuss additional constraints are presented. 
\subsection{Generative Adversarial Networks}
GANs are a category of \textit{generative} models that are trained in an \textit{adversarial} fashion lending it the name. The generative aspect of a GAN is through a generative network called the generator ($G$) while the adversarial aspect is through a discriminative network called the disciminator ($D$). In the original paper \citep{goodfellow2014generative}, the authors define $G$ as a function with parameters $\theta_{g}$ and construct a mapping $G(z;\theta_{g})$ where $z$ is a sampled from a noise distribution.  The discriminator $D$ has parameters $\theta_{d}$ and applies the mapping $D(x;\theta_{d})$ to an input $x$ providing a scalar output.$x$ comes from either the data distribution or the \textit{generated} distribution. The goal of a GAN is to train both $D$ and $G$ simultaneously such that $D$ learns to maximize the probability of identifying whether an input $x$ comes from data distribution or the \textit{generated} distribution while G tries to minimize this probability while generating samples. 
This is formally presented as,
\begin{equation}\label{eq:gan_equation}
        \begin{aligned}
        L_{adversarial}(G,D) &= \mathbb{E}_{x \sim  p_{data}}[log(D(x))] \\
        &  + \quad \mathbb{E }_{z  \sim  p_{generated}}  [log(1 - D(G(z)))]
    \end{aligned}
\end{equation}
Equation \ref{eq:gan_equation} shows the combined objective for the generator and discriminator where $p_{data}$ and $p_{generated}$ are the data and generated distributions respectively. The generator tries to minimize this objective while the discriminator tries to maximize it, which is formulated as,
\begin{equation}\label{eq:gan_equation2}
        \begin{aligned}
        G^{*} = \underset{G}{min} \underset{D}{max} L_{adversarial}(G,D) 
    \end{aligned}
\end{equation}
\subsection{Image-to-Image translation using GANs}
Generative-adversarial methods have also been extended for conditional data --- where a generated distribution is conditioned on additional information \citep{mirza2014conditional}. Several modes of information such as text, labels and images have been used as conditional information. Pix2pix \citep{Isola2016} uses images as input in a conditional fashion to generate translations of those images. Here the sampled noise $z$ is conditioned with an input $x$. In addition to the adversarial loss $L_{adversarial}$, a L1 loss between the generated and input conditional image is proposed. The combined loss is presented as,
\begin{equation}\label{eq:pix2pix_eq}
        \begin{aligned}
        L_{pix2pix}(G, D) &= L_{adversarial}(G, D) \\
        & + \quad \lambda \mathbb{E}_{x, y, z}[|| y - G(x, z) ||_{1}]
    \end{aligned}
\end{equation}
The authors interestingly show that the noise $z$ does not affect the GAN and can be completely eliminated leading to providing only $x$ sampled from the real data distribution as input. The concept of conditional GANs for image-to-image translation is extended to unpaired settings through the CycleGAN framework. 

\subsubsection{CycleGAN}
The CycleGAN architecture consists of two sets of generator and discriminator networks. Given images belonging to two domains $X$ and $Y$, the CycleGAN attempts to learn a mapping from $X \rightarrow Y$ through a network $G$. Discriminator $D_{Y}$ learns to differentiate if an image belongs to domain $Y$ or not and drives the training of $G$. The concept of cycle-consistency is enforced by learning the inverse mapping from $Y \rightarrow X$ through a network $F$. Similar to $D_{Y}$, $D_{X}$ exists for the inverse mapping. After mapping $X \rightarrow Y$ and $Y \rightarrow X$, the generated image is compared with the original by means of a cycle-consistency loss that ensures accurate reconstruction of the original image through the two mappings. 
\par
\begin{figure}[H]
    \centering
    \includegraphics[width=\textwidth]{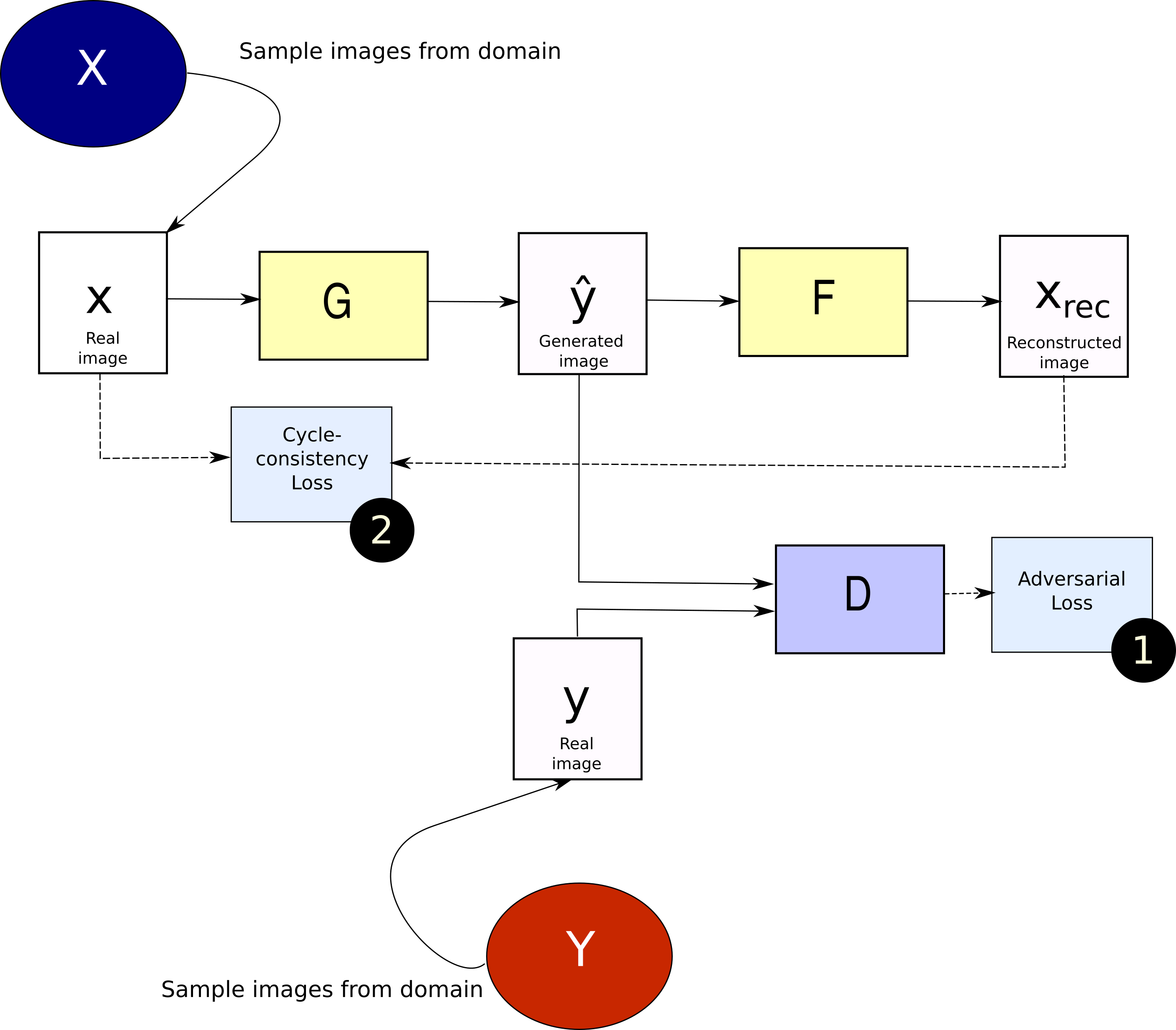}
    \caption{Block diagram of a CycleGAN architecture..}
    \label{fig:intro:cyclegan_arch}
\end{figure}

Figure \ref{fig:intro:cyclegan_arch} shows a diagram of the CycleGAN architecture for learning a mapping from domain $X \rightarrow Y$. The adversarial loss is similar to Equation \ref{eq:gan_equation} but with image $x$ from domain $X$ as input instead of a noise vector $z$. The cycle-consistency loss in the standard CycleGAN is a L1 loss between the input and reconstructed image. In the original paper \citep{Zhu2017}, the combined loss is formulated as, 
\begin{equation}\label{eq:cyclegan}
    \begin{aligned}
        L_{CycleGAN}(G, F, D_X, D_Y) &=  L_{adversarial}(G, D_{X}) +   L_{adversarial}(F, D_{Y})  \\
            & + \quad \lambda_A \mathbb{E}_{x \sim p_{data(x)}}[||F(G(x)) - x||_1] \\
            & + \quad \lambda_B \mathbb{E}_{y \sim p_{data(y)}}[||G(F(y)) - y||_1] 
    \end{aligned}
\end{equation}

\subsection{Constraints enforced on a CycleGAN}\label{sec:intro:constraints}
The authors of CycleGAN motivate the addition of a cycle-consistency loss by stating that there might be infinitely many mappings G that can produce an output distribution that is identified as belonging to the target domain. They also note that mode-collapse is quite common when only adversarial mechanics are used for training. They propose addition of structure to their training objective through a cycle-consistency loss as formulated in Equation \ref{eq:cyclegan}. The nature of the cycle-consistency constraint and other additional constraints that have been proposed in literature, are described in the following sections.
\subsubsection{Cycle consistency constraints}
Adding a cycle-consistency objective constraints optimization such that \\ ~$F(G(x)) \approx  x$ and $G(F(y)) \approx  y$. This objective is balanced against the adversarial objective through introduction of $\lambda_{A}$ and $\lambda_{B}$ as seen in Equation \ref{eq:cyclegan}. The balance between the adversarial objective and cycle-consistency objective can promote different characteristics during translation of the image. Increasing $\lambda$ can lead to stronger optimization of the cycle-consistency leading to better reconstructed images while the translated images might worsen as the adversarial objective tends to be weaker and vice-versa. Determining an ideal value of $\lambda$ generally involves a hyper-parameter search and can become unfeasible for training large, time consuming 3D networks as is the case with CBCT improvement. This begs the question -- \textit{can these $\lambda$ values be determined automatically? And can they be adapted during training by observing different training metrics and use them as proxy controls?} In the first part of the thesis, we try to answer these questions in a general manner that might be relevant across multiple tasks. 
\subsubsection{Structural constraints}
\begin{figure}[H]
    \centering
    \includegraphics[width=0.5\textwidth]{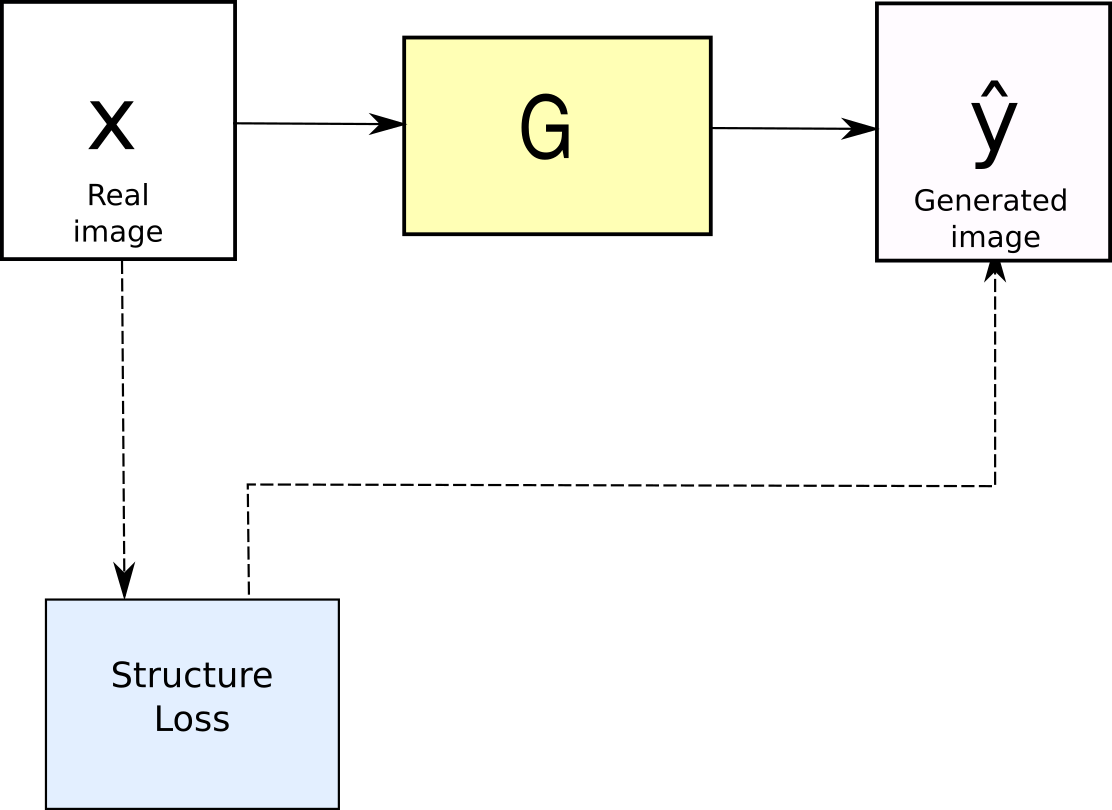}
    \caption{Design of structure loss for a CycleGAN architecture}
    \label{fig:intro:structure_loss}
\end{figure}

\cite{Zhu2017} conduct ablation studies across various datasets to demonstrate that stable performance is obtained when both cycle-consistency and adversarial constraints are retained. Additional structural constraints may be imposed on the CycleGAN such as constraints between input and translated images similar to \cite{shrivastava2017learning} where they term it as a regularization loss. Equation \ref{eq:cyclegan_structure} shows the modification made to the original CycleGAN objective shown in Equation \ref{eq:cyclegan} upon addition of the structure loss.
\begin{equation}\label{eq:cyclegan_structure}
    \begin{aligned}
       L_{CycleGAN+structure}(G, F, D_X, D_Y) &= L_{CycleGAN}(G, F, D_X, D_Y)  \\
        & + \quad \lambda_{struct}\mathbb{E}_{x \sim p_{data(x)}}[d(G(x), x)]  \\ 
        & + \quad \lambda_{struct} \mathbb{E}_{y \sim p_{data(y)}}[d(F(y), y)] \\
    \end{aligned}
\end{equation}
where $d(., .)$ is the distance metric or the specific loss function used between the images in source and target domain and $\lambda_{struct}$ is the weight associated with the structure loss.  Figure \ref{fig:intro:structure_loss} shows an extension of Figure \ref{fig:intro:cyclegan_arch} where the structure loss/ constraint is applied between the real image and generated image.\\

\cite{yang2018unpaired} also present a structure-constrained CycleGAN and show clear benefits of the additional structural constraints for the use-case of MRI to CT synthesis. Another research question posed as a part of the thesis is --- \textit{can we benefit through the addition of structural constraints to a CycleGAN? How do some of the existing methods, such as,  the MIND loss \citep{Heinrich2012} as presented in the work of \citep{yang2018unpaired} fare? Can simplified and potentially generalized losses also prove to be effective?} We aim to explore and provide answers to the questions posed above as the second part of the thesis. 

\section{Evaluation of CycleGAN-based methods}
Due to its unpaired nature, evaluation of CycleGAN-based translation poses quite a challenge. Fortunately, for the use-case of CBCT enhancement, we have the benefit of being able to construct some weak pairs consisting of CBCT images and their matching CT images. This process will be described in detail in Section \ref{methods:data_preprocessing}.  The set of weak pairs is split into validation and testing allowing evaluation of the translation during training as well as on unseen data post-training. Standard image similarity metrics are used to evaluate image translation quantitatively along with qualitative evaluation. Qualitative evaluation looks at multiple standpoints such as the introduction of artifacts, translation quality and resolution. 

\subsubsection{Out of Distribution Analysis}
Deep Learning approaches are susceptible to drastic performance drops when shown data that is out of the training distribution. In the case of generative approaches like the CycleGAN, presenting out of distribution data might induce unexpected artifacts during the translation putting into question the robustness of these methods. In order to test the generalizability of methods in this thesis, phantoms are used as out-of-distribution data. Phantoms are used to calibrate CBCT and CT machines and have different properties when compared to patient data and are described in Section \ref{methods:dataset:cbct}. Moreover, unlike weak pairs formed with patient data, phantoms can be used as perfectly matched pairs as their composition is static.  

\subsubsection{Domain-specific analysis}
Enhancement of CBCT scans can play a key role in adaptive radiotherapy, however, improvement in image similarity metrics might not solely show their validity for this purpose. Domain-specific analysis of the improved CBCT through comparing HU value distribution and line profiles with a matching CT image can provide more clinical relevant evaluation in this case. An automated lung segmentation model is also applied on the original and improved CBCT and compared with ground truth contours. Through these comparisons, we aim to show that improvement in CBCT quality benefits various down-stream applications of CBCT. 

\section{Clinical value and impact}
CBCT acquisition through onboard imaging during treatment delivery improves outcomes for treatment through more accurate positioning. Since acquiring this CBCT image is convenient and less intrusive compared to obtaining a full CT scan, improvements in quality of the CBCT, by making it CT-like can add a lot of value to the treatment process. For proton therapy, rescanned CTs are captured at certain intervals to reassess the treatment plan as proton-based plans are highly susceptible to changes observed during treatment. By having CBCTs similar to CT quality, rescanning may be reduced or avoided saving time, costs and improving patient convenience. For photon therapy, more accurate plans can be generated through these improved CBCT images and adapting treatment plans. However, clinical adoption of deep learning-based generative methods are challenging due to their black-box nature and inability to offer guarantees on new data. Evaluating methods on sufficiently large test sets, out of distribution analysis and qualitative inspection can offer more insight into their robustness. In addition to evaluation, deploying these models in clinical workflows with well described model cards \citep{Mitchell_2019} and data sheets describing the expected and potentially unexpected operational scenarios can help with their acceptance. Clinicians can then evaluate these independently on large cohorts of data and determine if they are suitable for clinical use. 
\section{Contributions}\label{sec:intro:contrib}
The thesis and its various investigations aim to answer the following research questions,

\begin{enumerate}
    \item Are default CycleGAN constraints sufficient to produce clinically viable enhanced CBCT images, if not, what are their failure modes?
    \item Where does one start adapting constraints? What constraints of the CycleGAN affect generated image quality the most?
    \item Can informed and data-driven constraints be developed for CBCT image enhancement?
    \item In order to evaluate image quality, what quantitative metrics can be leveraged effectively in such a semi-supervised setting?
\end{enumerate}

One of the research questions that proposed to investigate if constraints that work well for CBCT image enhancement generalize well to other medical imaging tasks was dropped in favour of deeper investigation of data-driven constraints and domain-specific evaluation. As future work, another medical image dataset will be incorporated to demonstrate generalizability of our methods.

\chapter{Methods and Materials}
\section{Dataset}
This section describes the data used to answer the questions this thesis proposes. Two datasets, one comprising of natural images and the other of medical images, are considered. Natural image dataset allows for quick experimentation and also helps demonstrate if the methods explored in the thesis transfer between natural and medical images. 

\subsection{\textit{Map $\leftrightarrow$ aerial photo} Dataset}

\begin{figure}[H]
    \centering
           \includegraphics[width=\textwidth]{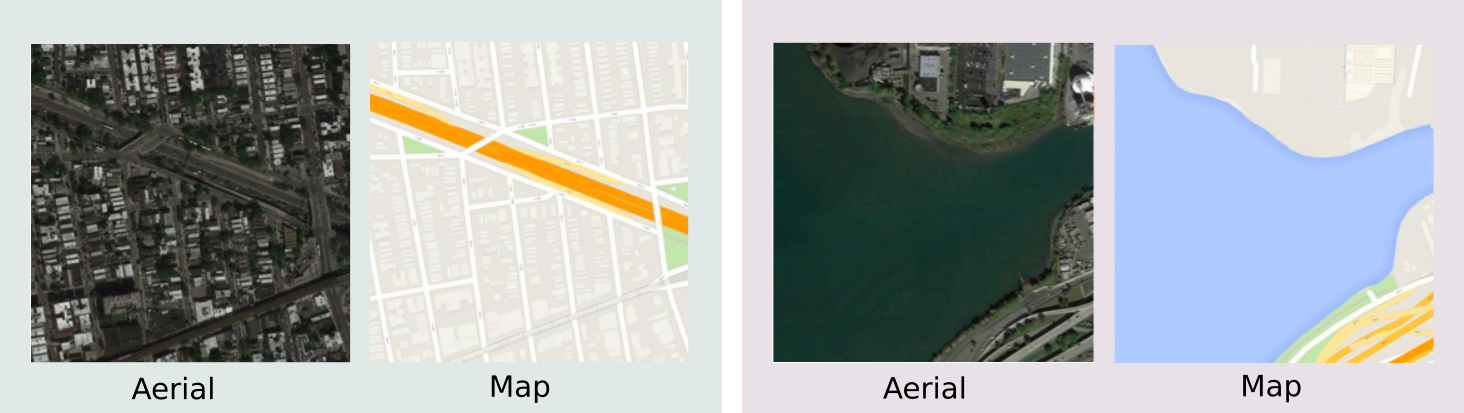}
     \caption{Two pairs of matching images for the \textit{map $\leftrightarrow$ aerial photo} dataset}
    \label{fig:dataset:maps}
\end{figure}

The \textit{map $\leftrightarrow$ aerial photo} dataset was introduced by~\cite{Isola2016} in their pix2pix paper and contains 2194 images scraped from Google Maps. Since map data and satellite data have one-to-one correspondence, they were used in a paired form in the original paper. \cite{Zhu2017} use this dataset in an unpaired form to evaluate the CycleGAN and additionally, compare it against the pix2pix. Figure \ref{fig:dataset:maps} shows two sample pairs from the dataset. The \textit{map $\leftrightarrow$ aerial photo} dataset was chosen as it contains certain geometrical structures that are preserved between source and target domains. Structures such as clusters of houses, road/highway patterns, and boundaries between land and water are retained across both domains. In contrast to the original paper, we use \textit{aerial} as the source domain and \textit{map} as the target domain in order to be able to better evaluate the generated images as map images are visually simpler.

\subsection{Radiotherapy treatment dataset with CBCT and CT images}
\label{methods:dataset:cbct}
\begin{figure}[H]
    \centering
    \includegraphics[width=\textwidth]{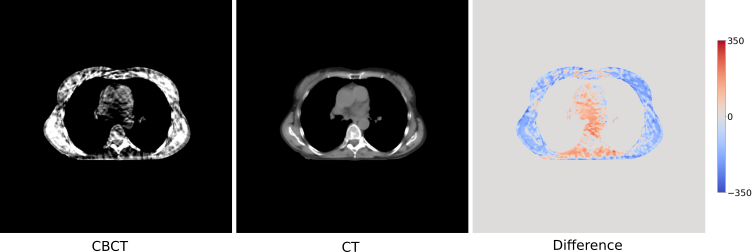}
    \caption{Mid-axial views of the CBCT, CT and their difference image for a patient. The difference image is shown in the HU scale capped between -350 and 350. }
    \label{fig:methods:cbct_ct}
\end{figure}

The dataset chosen for the use-case of CBCT enhancement, comprises of CBCT and CT scans acquired during proton beam radiation therapy. Seventy-two lung cancer patients were considered for treatment with the treatment delivery conducted over multiple fractions. At each fraction, a CBCT image was captured resulting in a total of 774 CBCT scans across all patients. In addition to planning CTs, rescanned CTs are also collected to verify/adapt treatment plans, leading to a total of 257 CT scans. Figure \ref{fig:intro:cbct_ct_examples} shows the CBCT and CT scan from a randomly selected patient. The CBCT images are acquired through Mevion CBCT scanners which are suspectible to a significant amount of noise (sources of which are highlighted in Section \ref{sec:intro:improvement}) while reconstructing the images. Figure \ref{fig:methods:cbct_ct} shows images of the CBCT and CT using a soft-tissue window that helps focus on values within the heart, skeletal muscle, etc. Compared to the CT, the CBCT image has very different intensity values (bad HU calibration) along with streaks in the image (due to scattering and motion artefacts). Table \ref{tab:methods:datasheet} shows the datasheet for this dataset filled with information that is made available to us. It is designed in such a manner that it should allow a layperson to understand more about the dataset and its nature. 

\begin{table}
  \noindent\makebox[\textwidth]{%
  \resizebox{1.5\textwidth}{!}{

\begin{tabular}{ll}
\hline
\multicolumn{2}{c}{\textbf{CBCT-CT Dataset}}                                                                                                                                                                                                                                                                                                                                                                                     \\ \hline \vspace{2mm}
For what purpose was the dataset created?                                                                                                                                                                  & Radiation therapy treatment                                                                                                                                                                                           \\ \vspace{2mm}
\begin{tabular}[c]{@{}l@{}}Who created the dataset and on\\ behalf of which entity\end{tabular}                                                                                                            & Maastro Clinic                                                                                                                                                                                                        \\ \vspace{2mm}
What do the instances that comprise the dataset represent?                                                                                                                                                 & Patients undergone proton-beam radiotherapy                                                                                                                                                                           \\ \vspace{2mm}
How many instances are there in total?                                                                                                                                                                     & 72                                                                                                                                                                                                             \\ \vspace{2mm}
\begin{tabular}[c]{@{}l@{}}Does the dataset contain all possible instances or is it a sample \\ of instances from a larger set?\end{tabular}                                                               & Contains all possible instances from the treatment process                                                                                                                                                            \\ \vspace{2mm}
What data does each instance consist of?                                                                                                                                                                   & \begin{tabular}[c]{@{}l@{}}Each instance consists of multiple CBCTs used for positioning \\ and multiple repeat CTs used to recalculate the treatment plan\end{tabular}                                               \\ \vspace{2mm}
Is there a label or target associated with each instance?                                                                                                                                                  & \begin{tabular}[c]{@{}l@{}}Each instance consists of annotated organs and tumours that could be used \\ as targets for segmentation\end{tabular}                                                                      \\ \vspace{2mm}
Are relationships between individual instances made explicit?                                                                                                                                              & Yes. No relationships exist between the patients                                                                                                                                                                      \\ \vspace{2mm}
Are there recommended data splits?                                                                                                                                                                         &                                                                                                                                                                                                                       \\ \vspace{2mm}
\begin{tabular}[c]{@{}l@{}}Are there any errors, sources of noise, or redundancies in the\\ dataset?\end{tabular}                                                                                          & Redudancies in the form of multiple CT scans per patient might exist                                                                                                                                                  \\ \vspace{2mm}
\begin{tabular}[c]{@{}l@{}}Is the dataset self-contained, or does it link to or otherwise rely on\\ external resources\end{tabular}                                                                        & Self-contained                                                                                                                                                                                                        \\ \vspace{2mm}
Does the dataset contain data that might be considered confidential                                                                                                                                        & Yes, senstive patient data                                                                                                                                                                                            \\ \vspace{2mm}
\begin{tabular}[c]{@{}l@{}}Does the dataset contain data that, if viewed directly,\\  might be offensive, insulting, threatening,\\  or might otherwise cause anxiety?\end{tabular}                        & Yes, dataset consists of patients with cancer                                                                                                                                                                         \\ \vspace{2mm}
Does the dataset relate to people?                                                                                                                                                                         & Yes                                                                                                                                                                                                                   \\ \vspace{2mm}
Does the dataset identify any subpopulations?                                                                                                                                                              & \begin{tabular}[c]{@{}l@{}}Not directly, but information from the scans\\  may be used to identify gender, age, weight etc.\end{tabular}                                                                              \\ \vspace{2mm}
\begin{tabular}[c]{@{}l@{}}Does the dataset contain data that might be considered sensitive\\ in any way?\end{tabular}                                                                                     & Yes                                                                                                                                                                                                                   \\ \vspace{2mm}
How was the data associated with each instance acquired?                                                                                                                                                   & \begin{tabular}[c]{@{}l@{}}For each instance, CBCTs were acquired using Mevion CBCT scanners \\ and CTs were acquired using Philips Sensation 16 CT scanner\end{tabular}                                              \\ \vspace{2mm}
Were any ethical review processes conducted?                                                                                                                                                               & Approval through an Institutional Review Board before internal access to data                                                                                                                                         \\ \vspace{2mm}
Was any preprocessing/cleaning/labeling of the data done?                                                                                                                                                  & Not in its original form                                                                                                                                                                                              \\ \vspace{2mm}
Has the dataset been used for any tasks already?                                                                                                                                                           & \begin{tabular}[c]{@{}l@{}}Yes, for treatment planning. \\ In terms of deep learning based approaches, \\ - Used for CUT implementation\\ - Comparison between different models for unpaired translation\end{tabular} \\ \vspace{2mm}
\begin{tabular}[c]{@{}l@{}}Will the dataset be distributed to third parties outside of the entity \\ (e.g., company, institution, organization) on behalf of which\\ the dataset was created?\end{tabular} & No, it is currently an internal Maastro dataset                                                                                                                                                                       \\ \hline
\end{tabular}
}}
\caption{Datasheet for the CBCT-CT dataset with relevant information to allow easier understanding as per \cite{gebru2020datasheets}.}
\label{tab:methods:datasheet}
\end{table}

\subsubsection{Phantoms}\label{sec:methods:phantoms}
\begin{figure}[H]
    \centering
    \includegraphics[width=\textwidth]{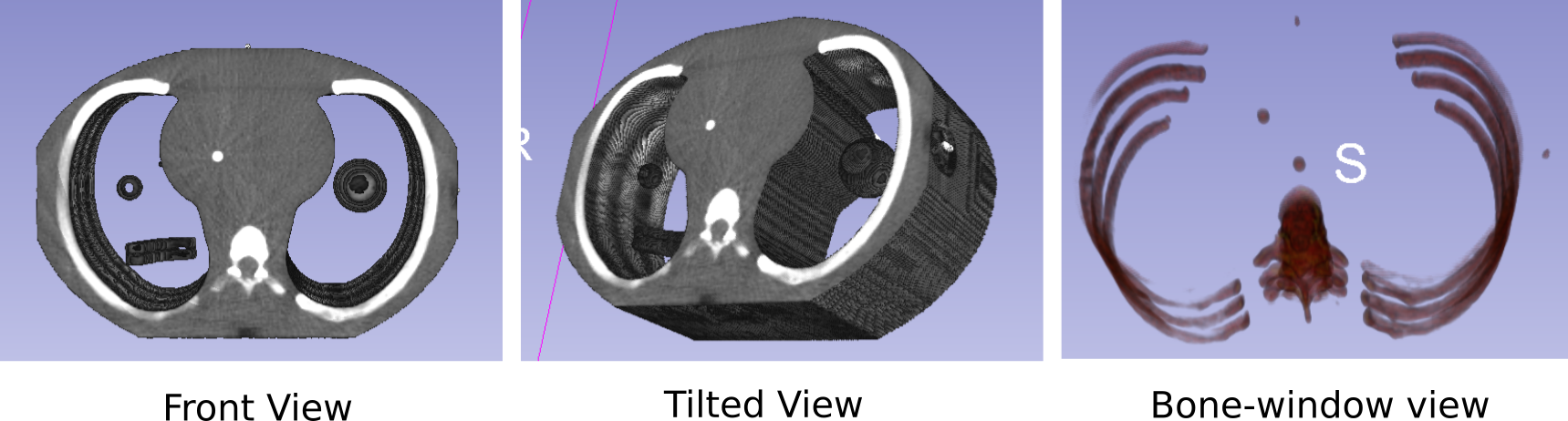}
    \caption{Anthropomorphic phantom shown with front and tilted views. A bone-window view is also shown where intensities are windowed to expose only the bones. As can be seen, the phantom replicates a human-like anatomy.}
    \label{fig:methods:phantom}
\end{figure}

Imaging phantoms are objects designed with a known geometrical and physical composition which are used for quality assurance and evaluation of CT machines \citep{Medicine1977}. In addition to the patient data, we also have access to phantoms scanned on the same CBCT and CT machines as the patients. Due to their known composition, phantoms can be used as true ground truth pairs, termed as \textit{strong pairs}, in contrast to weak pairs that were mentioned earlier. This is because weak pairs can contain differences due to motion, setup and biological changes while phantoms are not subject to these changes. Figure \ref{fig:methods:phantom} shows the anthropomorphic phantom available as a part of the dataset. 

\section{Data Pre-processing}\label{methods:data_preprocessing}
Data pre-processing is a key element in the machine/deep learning pipeline converting existing representations to representations that are best suited for optimization. This is especially important for medical images where inconsistent representations of data can lead to biases that propagate through the modelling process \citep{Raad2021}. In addition to preprocessing the data, the splits used for training, validation and test for each of the datasets are also mentioned. 
\subsection{\textit{Map $\leftrightarrow$ aerial photo } Dataset}
The dataset is randomly split into 1085, 9 and 1098 images for train, validation and test, respectively. A small number of images are chosen for validation to allow extensive manual inspection of translation properties. During training, the dataset is resized to $256\times256$, and augmented with random horizontal flipping of the images. The data is then normalized between $(-1, 1)$. For validation and testing, only resizing and normalization are performed.

\subsection{CBCT and CT Scans}
The general standard for capture and communication of medical images, is the DICOM (Digital Imaging and Communications in Medicine) format. DICOM is a rich format providing extensive metadata about the captured images in the form of sets of data that contain various attributes within. In addition to the storage of CT and CBCT scans, the format also allows storage of annotated contours, treatment plans and dose distributions. All the data used for this thesis was made available in the DICOM format. Multiple steps of data preparation and processing need to be undertaken to convert the scans from DICOM to images that can be fed to deep learning models. This section highlights steps that are taken for processing the data starting from format conversion and resampling to split specific processing. 

\subsubsection{Data Conversion, Resampling and Stratification}
The DICOM files for CBCT and CT scans are first converted to an NRRD (Nearly Raw Raster Data) format. The NRRD format retains only lightweight metadata that is needed to determine the orientation and spacing of the 3D image in a global coordinate system. Patient, scanner and process-specific information is thus, discarded. DICOM stores annotated contours as points on a plane in a format called RTSTRUCT (RT Structure Sets). Along with the CBCT/CT scans, the RTSTRUCTS are converted to binary labelmaps stored in NRRD format, as well. \par
\begin{figure}[H]
    \centering
    \includegraphics[width=\textwidth]{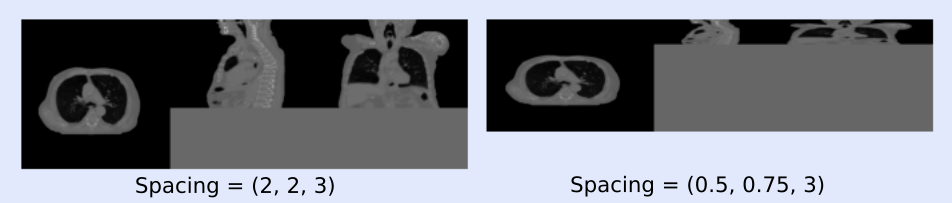}
    \caption{Difference in images rendered by an imaging software that does not account for pixel spacing. The same scan appears very different for the two spacings. For the spacings, the first term denotes the x-axis, second term the y-axis and third term the z-axis. Note that the pixel sizes of the images are different with the first and second images having a size of $259\times259\times59$ and $1003\times670\times59$ respectively.}
    \label{fig:methods:spacing}
\end{figure}

The CT scans in the dataset are acquired with different field-of-view at the discretion of the radiation therapist. Due to constant size of the line detector, increase in field-of-view leads to a larger pixel sizes and therefore, increased spacing between them. This needs to be taken care of during pre-processing, as without this the image representations can be misleading. Figure \ref{fig:methods:spacing} shows an example of the same CT scan with different spacings. 
In order to minimize interpolation, the resampling is done as follows,
\begin{enumerate}
    \item Get frequency counts of spacing of all scans in the dataset.
    \item Sort spacings in ascending order and rank all spacings based on their frequency counts. 
    \item Select the smallest rank starting from bottom of the list.
\end{enumerate}
On following the procedure above, $1.2695\times1.2695\times3$ is determined as the ideal spacing. The CBCT scans are also resampled to the resolution of the CT as we expect the CBCT to be a replacement for the CT and hence, possess similar resolution. All of the above are performed using the \textit{SimpleITK} library \citep{Bradley2013} in python. \par
The dataset is stratified at the patient-level with 50 patients in the train set, 4 in the validation set and 18 in the test set. Do note that due to availability of more than a single pair per patient, the instances available for train and val are larger than the number of patients. These are specifically mentioned in the pre-processing subsections. 

\subsubsection{Train Set Pre-processing}
In order to leverage the full benefits of unpaired data, training is done by selecting a random CBCT and CT scan from all the available set of scans. Since the selection is completely random, a CT scan from one patient can be paired with a CBCT scan from the same patient or another patient. We hypothesize that this allows for learning more generalized properties and also allows balancing training instances in cases where a particular patient may have fewer CT or CBCT scans. A total of 736 CBCT scans and 219 CT scans, from the 50 patients, are used during training. A set of \textit{online } pre-processing steps are followed during training , as shown in Algorithm \ref{alg:train:preprocessing}.

\begin{algorithm}[H]
\caption{Steps for processing CT/CBCT data in an online manner during training}
\label{alg:train:preprocessing}
\Function{process\_CT\_CBCT(CTpath, CBCTpath):}{
    \Comment{Load CT and CBCT from NRRD files}
    CT, CBCT = load\_image(CTpath), load\_image(CBCTpath)\;
    
    \Comment{Generate body mask and apply it to the CT and CBCT images}
    CT, CBCT = apply\_body\_mask(CT), apply\_body\_mask(CBCT)\;
    
    \Comment{Limit the axial range of the CT to match the CBCT}
    CT = truncate\_CT(CT, CBCT)\;
    
    \Comment{Sample patches of a given size from the CT and CBCT}    
    CT\_patch, CBCT\_patch = get\_patch\_pair(CT, CBCT);
    
    \Return{CT\_patch, CBCT\_patch}
}
\end{algorithm}

Each of the functions in the pre-processing steps are described below,
\begin{itemize}
    \item \texttt{\bfseries apply\_body\_mask}: A body mask based on HU value thresholding is generated for both the CBCT and CT scans. This is done in order to minimize the effect of out-of-body structures during training. The body masking algorithm takes in a CBCT/CT scan and binarizes the image based on a threshold. After binarization, for each slice in the 3D scan, the largest contour is found. This contour is filled and finally a smoothing kernel is applied over the contour points in order to have smooth masks. 
    \item \texttt{\bfseries truncate\_CT}: Truncation of the CT aims to limit the field-of-view (FOV represents the amount of physical space captured in the image) of the CT scan in the z-axis. This is done so that the FOV of the CBCT and CT are roughly matching and correspond to similar anatomical regions. A registration is performed using \textit{SimpleITK} where CBCT is the fixed image and CT is the moving image. The resulting transformation is used to truncate the CT scan at matching coordinate locations of the CBCT scan. 
    \item \texttt{\bfseries get\_patch\_pair}: Loading full volumes during training is limited by memory size and computational time requirements. To tackle this, training is usually done in a patch-based manner. This function takes in a CBCT and CT scan and samples patches from random locations in the CT. For CBCT, the CT location is used as a starting point with small perturbations added to allow more randomness during training. 
\end{itemize}

\subsubsection{Validation and Test Set Pre-processing}
For each of the patients in the validation and test set, multiple CBCTs and rescanned CTs are available that are acquired through the treatment process. Rescanned CTs acquired at a time point close to the CBCT generally have good anatomical correspondences although there may be differences due to setup and random errors. These correspondences can be leveraged in order to form weak pairs, which can then be used to evaluate translation quantitatively. The process followed to generate these weak pairs is highlighted below,
\begin{enumerate}
    \item Select the rescanned CT and the CBCT closest to it. The maximum time-difference between the two is limited to one day, so that scans with potentially larger anatomical changes are ignored.
    \item The rescanned CT is registered to the CBCT through deformable registration using parameters from the \textit{SimpleElastix} library \citep{Kasper2016}. 
    \item Apply the registration transform to the rescanned CT and available contours (only available on the test set). 
\end{enumerate}

\begin{figure}[H]
    \centering
    \includegraphics[width=0.8\textwidth]{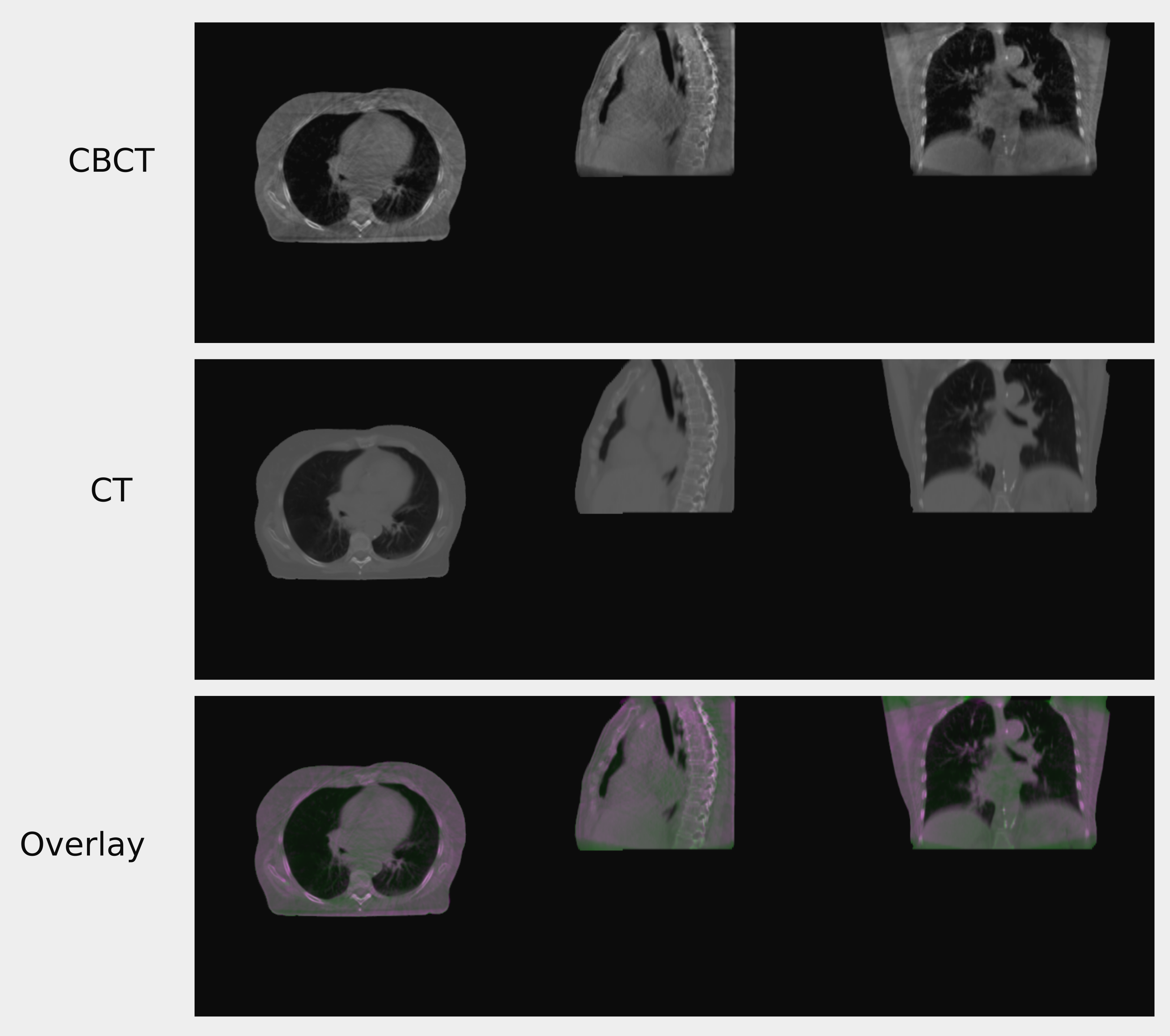}
    \caption{Registered CBCT, CT and overlay image. In the overlay image, the green and pink areas in the image show differences between the CT and CBCT. Green shows values that are greater in the CT while pink shows values that are greater in the CBCT. }
    \label{fig:methods:registered}
\end{figure}

A sample of the weak pair formed from the process above, is shown in Figure \ref{fig:methods:registered}.  Out of the four patients in the validation set, 20 pairs are formed that are used for evaluating model checkpoints. The 18 patients in the test set are chosen more restrictively by manually ensuring matches between the CBCT and rescanned CTs, and therefore, contain only 18 pairs used during testing. The manual inspection is done by medical physicists at Maastro Clinic. 

\section{Adaptive Control of CycleGAN Constraints}\label{sec:methods:adaptive_control}
The constraints of a CycleGAN that aid in its translation, namely, adversarial and cycle-consistency loss, are generally balanced statically. Well-performing values for $\lambda$ (both $\lambda_A$ and $\lambda_B$) can generally be chosen through hyper-parameter search. However, this is resource-intensive and may not transfer across different datasets and model configurations. Furthermore, training may benefit by using more dynamic balancing strategies, similar to dynamic augmentation proposed by \cite{karras2020training}. In the paper, \cite{karras2020training} propose using discriminator outputs as a \textit{control signal} to determine if augmentation would benefit the training. If these \textit{control signals} capture training dynamics effectively, then they could also prove useful for a dynamic adaptation of constraints. Thus, inspiration from this work was taken to develop the following \textit{control signals} in an adaptive form of the CycleGAN called AdaptiveCycleGAN,

\subsubsection{Discriminator-based Control Signals}
The discriminator-based control signal is very similar to the one used by \cite{karras2020training}. It looks at the outputs of the discriminator and formulates the parameter, $r_t$. 
\begin{equation}\label{eq:methods:rt}
    r_t = \mathbb{E}[sign(D_{train})]
\end{equation}
where $D_{train}$ represents the discriminator output and the expectation is computed over four mini-batches of 64 images each. If $r_t > 0.6$, then the adaptive augmentation strategy is triggered. \par
In the case of the CycleGAN, the discriminator is PatchGAN-based and outputs a 3D patch. To account for that, Equation \ref{eq:methods:rt} is modified such that $D_{train}$ corresponds to the mean value over all voxels of the output patch. After some initial experiments, it was determined that $D_{generated}$ is more responsive to training dynamics than $D_{train}$ and was used henceforth. The number of images that $r_{t}$ is computed over is set to 256, although, the number of mini-batches vary based on the batch-size selected. Finally, the adaptive lambda strategy is triggered at certain frequency intervals, as it was seen with preliminary experiments that adapting too often leads to early divergence.

When the adaptive $\lambda$  strategy is triggered, the $\lambda$  values are changed by a percentage of 0.05. It is increased if $r_t > 0.6$ as the discriminator is largely predicting generated images as real. This generally means that the generator is overpowering the discriminator and can focus less on the adversarial loss. Similarly $\lambda$ is decreased when $r_t < 0.6$. 

\subsubsection{Cycle-consistency-based Control Signals}
The intuition for cycle-consistency-based control signals is derived from various initial experiments that were run to understand training dynamics, interpretation of loss curves and their effect on the translations. The following observations were made, (1) for well-performing configurations, the cycle-consistency loss decayed smoothly while the adversarial loss had a high variance but with a mostly constant mean value; and (2) decaying cycle-consistency loss was generally a good indicator of strong reconstructions while high-variance adversarial loss was a good indicator of being able to capture the domain style. Based on this, the cycle-consistency loss curves are used as a \textit{control signal} for adaptive $\lambda$ strategy. 
Cycle-consistency-based control signal starts monitoring the cycle-consistency loss after a certain number of iterations of training have completed, termed generally as warmup iterations. Following this, a patience-based strategy similar to the one used in learning rate schedulers is leveraged, where, at a certain iteration a set of previous samples are retained and compared with the current value. If the current value is greater than all of the samples, then the adaptive $\lambda$ strategy is triggered. 

The adaptive $\lambda$ strategy for this signal, operates in two modes, threshold-based and relative. The threshold-based version increases/decreases $\lambda$ by a constant change rate while the relative version changes it by a rate that is equal to the relative difference between the current and the minimum value among previous samples. 

\section{Structure-consistency Losses}\label{sec:methods:structure_losses}
The second part of the thesis investigates methods that add additional constraints to the CycleGAN in the form of structure loss. As described in Figure \ref{fig:intro:structure_loss} and Section \ref{sec:intro:constraints}, the structure loss is added between the real image and the generated image in order to restrict the set of possible mappings between the source and target domains. This is not a new concept in the CycleGAN framework but its investigation has been fairly limited, especially for medical images. A challenge posed in the wider applicability of such structural losses, is that these tend to be domain-specific or rely on hand-crafted functions, while one of the primary benefits of the CycleGAN is that it is able to perform well in general as well as domain-specific settings without much configuration required. To address this problem, a generalized frequency loss, inspired by \cite{jiang2020focal}, is implemented that requires no configuration on the part of a user. However, to reap its full benefits, a preliminary check is advised to see if the loss would benefit a particular use-case. This is done through implementation of a script that uses a pre-existing dataloader and provides a report. Before the generalized frequency loss is described, some existing structure losses are described and subsequently investigated.

\subsection{Modality Independent Neighbourhood Descriptor \\ (MIND) Loss}
The MIND loss was used for MR-to-CT synthesis as a part of a structure-constrained CycleGAN \citep{yang2018unpaired}. Images from source and target domain are mapped to a common feature domain, which is done using the MIND descriptor. The MIND descriptor was originally developed for MR to CT registration and is therefore capable of handling cases where images are of different modalities. The MIND feature for an image $I$ at voxel $x$ is formulated as,
\begin{equation}
    F_x^{(\alpha)}(I) = \frac{1}{Z} exp(-\frac{D_P(I,x,\alpha)}{V(I,x)})
\end{equation}
where $\alpha$ is such that $x + \alpha$ is within a region called the non-local region. $Z$ is a normalization constant and $D_P$ is formulated as 
\begin{equation}
D_P(I, x, \alpha) = C * (I - I'(\alpha))^2
\end{equation}
\begin{equation}
V(I, x) =  \frac{1}{4}\sum_{n\in N} D_p(I, x, n)
\end{equation}
where $I'(\alpha)$ represents the image $I$ translated by $\alpha$; $C$ is a gaussian kernel with $\sigma=2$; and $N$ is the four-neighbourhood of voxel $x$. For each value of $\alpha$, a scalar value is computed resulting in a vector $F_x$, with 81 elements. The number of elements is given by the size of the non-local region which is $9\times9$. The MIND descriptor is used as MIND loss through the formulation below,
\begin{equation}
    d(G(x), x) = \frac{1}{N}\sum_{x} ||F_{x}(G(x)) -  F_{x}(x)||_{1}
\end{equation}\label{eq:methods:mind_loss}
where $G(x)$ is the predicted image by the generator, and $x$ is the real image. The difference in MIND descriptor representations is summed over all voxels in the images. $d(G(x), x)$ is used in Equation \ref{eq:cyclegan_structure} to formulate the full structure loss. For more details, the original paper by \cite{yang2018unpaired} can be referred to. 

\subsection{Registration Metric Losses}\label{sec:methods:reg_loss}
Since the MIND loss was originally based on a registration similarity metric, a less computation and memory intensive metric was tested in this thesis. The metric chosen was the Local Normalized Cross Correlation (LNCC) which is used by the \textit{DeepReg} toolkit. An implementation by the \textit{Monai} library \citep{Li2021} was chosen as it operates directly on pytorch tensors and therefore, is more efficient. Similar to the MIND loss,  the loss values are interpretable on the image space and can allow easy debugging. \\
\par
\begin{figure}[H]
    \centering
    \includegraphics[width=0.9\textwidth]{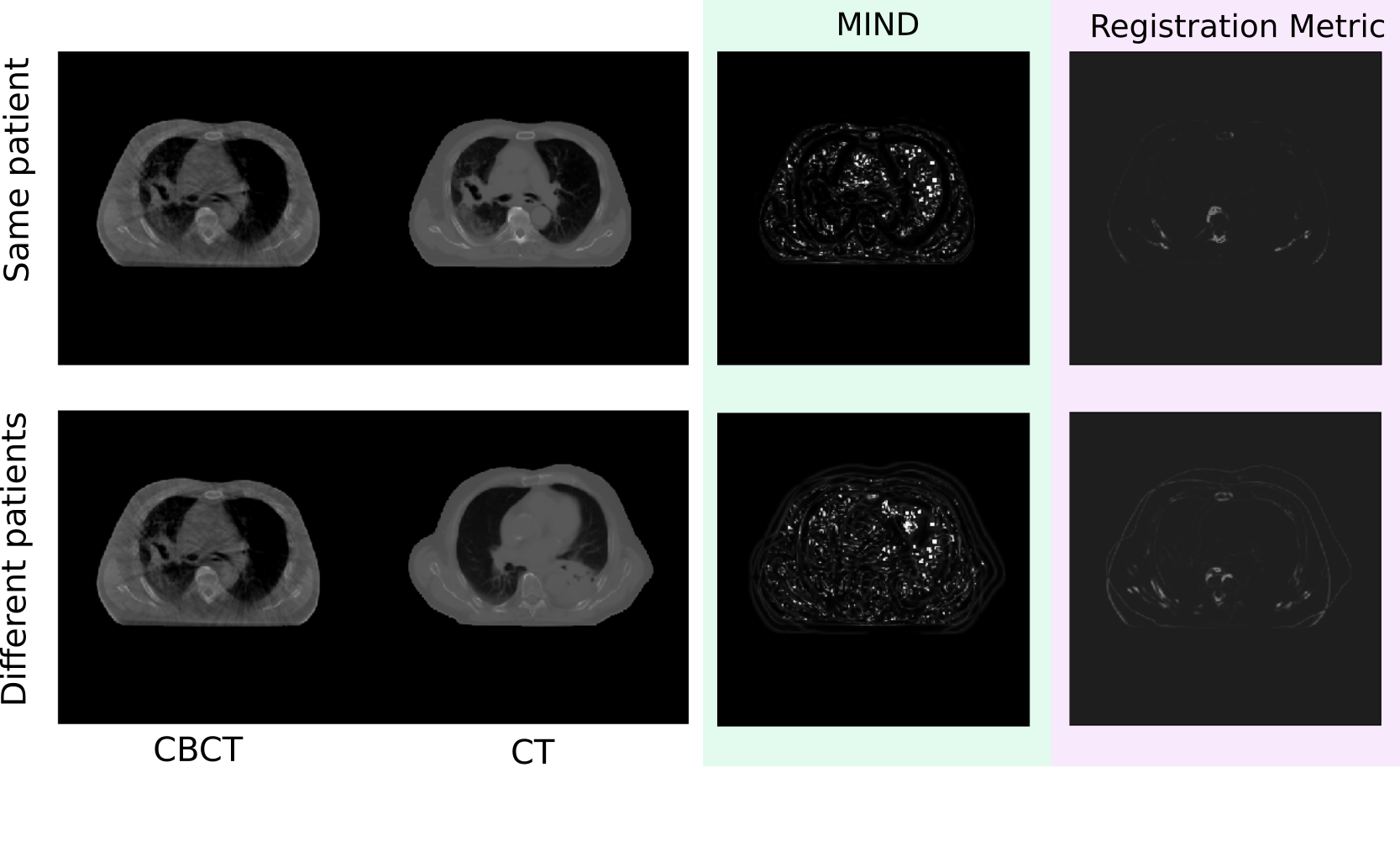}
    \caption{Absolute differences in MIND (shown in green) and registration metric (shown in red) representations between CBCT and CT images. The differences are shown on the mid-axial slice for CBCT and CT scans taken from the same patient and CBCT and CT scans taken from different patients.}
    \label{fig:methods:mind_reg_viz}
\end{figure}

Figure \ref{fig:methods:mind_reg_viz} shows the MIND and registration metric in the image space for mid-axial slices for the same patient as well as for different patients. Larger values are expected for different patients as their anatomies differ. The losses are also expected penalize deviations in anatomy when used during training. Registration metric shows larger values for different patients in the bony areas and body contours. For the MIND representation, it is seen that the intensities are not directly correlated with anatomical properties. 

\subsection{Generalized Frequency Loss}
Frequency spectrum representations of images can allow capturing patterns within the image, that may not be easy to identify in their spatial domain representations. In its frequency spectrum, the image is broken down into its sinusoidal components that can then be further analyzed by looking at their amplitudes and phase. Converting an image to its frequency spectrum representation involves a three-dimensional DFT transform, which is stated as follows,
\begin{equation}
    F(u, v, w) =  \sum_{x=0}^{L-1} \sum_{y=0}^{M-1} \sum_{z=0}^{N-1} I(x,y,z)e^{-j2\pi (ux/L+vy/M+wz/N)}
\end{equation}
where $L$, $M$ and $N$ are dimensions of the CT and CBCT scans. For the development of the generalized frequency loss, the frequency representation is then normalized as below,

\begin{equation}
    F^{'}(u, v, w) =  \frac{1}{\sqrt{LMN}}F(u, v, w)
\end{equation}
where $F^{'}(u, v, w)$ is then shifted such that zero frequency lies at the center of the image. Following the shift, only magnitude spectrum of the frequency spectrum is taken as $F_{mag}(u, v, w) = |F^{'}(u, v, w)|$. On the magnitude spectrum, a \textit{tanh} non-linearity is applied as,
\begin{equation}
    F_{rep}(u, v, w) = tanh(F_{mag}(u, v, w))
\end{equation}
The $tanh$ non-linearity is applied in order to scale all values to the range of 0 to 1. This was done to combat the differences in scale of frequency domain representations across different image sets. Alternatively, a careful strategy to ensure that images in the dataset are in similar scales while generating their frequency representations can be designed. However, this becomes extremely data-specific, and by no means, is generalized. Another more general solution is, computation of a log transformed magnitude spectrum, but this led to very different scales across different images tested. This would be unsuitable for direct comparison in the frequency domain between the two images. 

Addition of the $tanh$ makes several assumptions about the importance of different intensities in the magnitude spectrum, as it leads to the following effects: (1) values greater than $0.5$ are subdued and (2) as values increase in intensity, their rate of change is also dampened. However, the hypothesis is that the distribution of values, in the magnitude spectrum and not the intensities itself, are of primary importance. Note that intensity of values are still captured but only higher intensities are dampened. Due to ortho-normalization of the FFT, the values are not generally large to begin with. 

\begin{figure}[H]
    \centering
    \includegraphics[width=\textwidth]{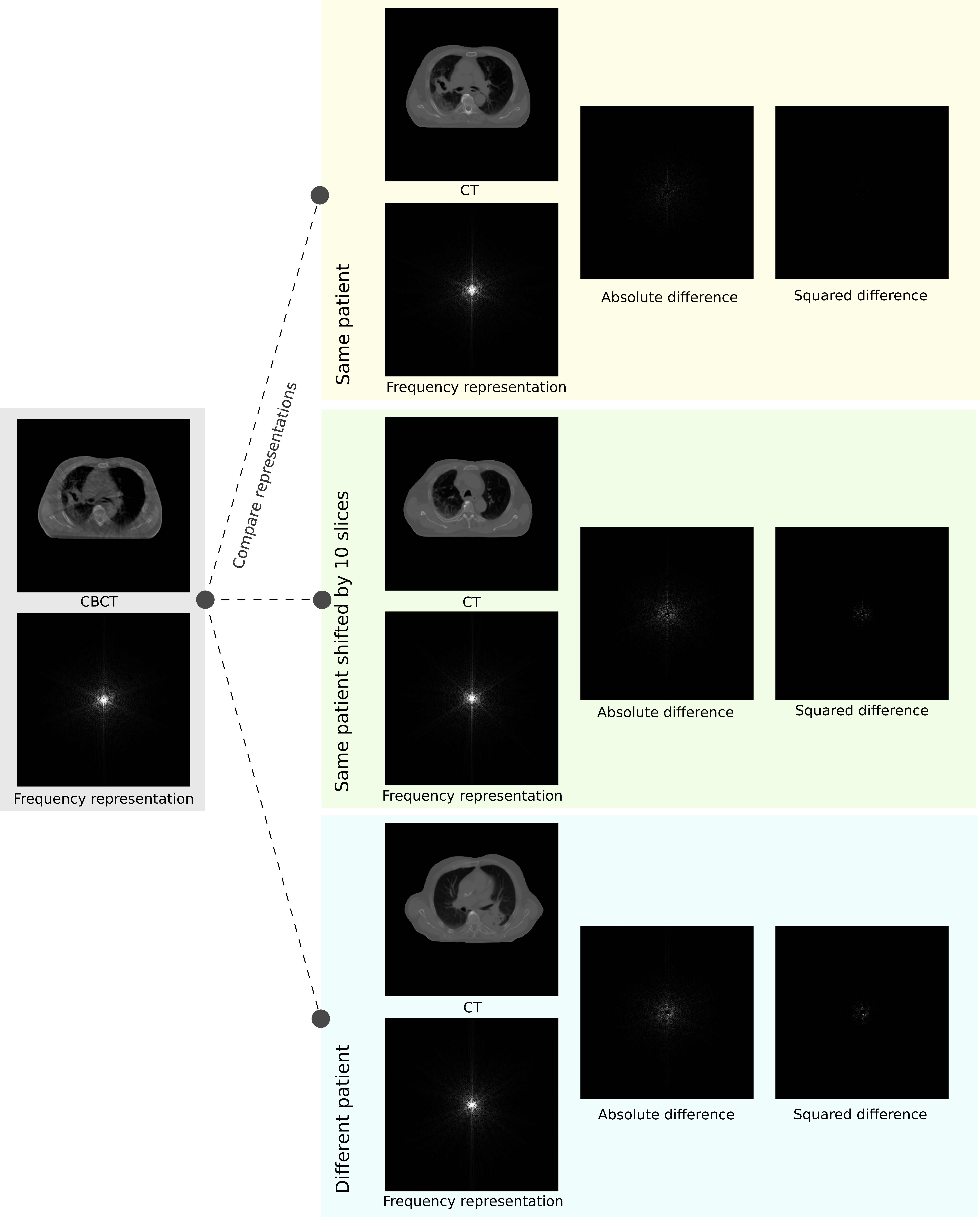}
    \caption{Generalized frequency representation of CBCT scan taken from one patient compared against CT scan taken from the following, (1) same patient, (2) same patient but offset by 10 slices and; (3) a different patient. The absolute and squared differences between representations are shown. }
    \label{fig:methods:freq_viz}
\end{figure}

Through detailed analysis on both the datasets used in the thesis, it was seen that the generalized representation captured differences in domains appropriately. A small difference in representation was seen for corresponding images of different domains, while a larger difference was seen for non-corresponding images. Figure \ref{fig:methods:freq_viz} shows the generalized frequency representations along with absolute and squared difference between these representations for different sets of images. 

Upon obtaining the generalized frequency representation, the generalized frequency loss is represented as,
\begin{equation}
    d(G(x), x) = \sum_{u, v, w} ||F_{rep}(u, v, w)(x) -  F_{rep}(u, v, w)(G(x))||_{1}
\end{equation}\label{eq:methods:gen_freq_loss}
where $G(x)$ is the predicted image by the generator, and $x$ is the real image. The difference in frequency representations is summed over all voxels in the images. $d(G(x), x)$ is used in Equation \ref{eq:cyclegan_structure} to formulate the full structure loss. 
The methods and evaluation implemented as a part of this thesis was built upon the \textit{ganslate} \citep{ibrahim_hadzic_2021_5494572} codebase.

\section{Evaluation}
Evaluation of unpaired translation methods is a non-trivial task and generally relies on a combination of quantitative and qualitative criteria. Domain-specific evaluation can also often be leveraged in order to determine if the generated images are suitable for downstream tasks. 

\subsection{Image Similarity Metrics}
Image similarity metrics common in image translation and quantitative IQA (Image Quality Assessment) such as Mean Absolute Error (MAE), Mean Squared Error (MSE), Normalized Mean Squared Error (NMSE), Power-to-Signal Noise Ratio (PSNR) and Structural Similarity Index Measure (SSIM) are used to quantitatively evaluate different methods. These metrics are outlined below,
\begin{itemize}
    \item Mean Absolute Error (MAE)
    \begin{equation}
        MAE(ref, pred) = \frac{1}{N} \sum_{i=1}^{N} | ref(i) - pred(i) |
    \end{equation}
    where $N= $total number of voxels in the image. $ref$, in our work, is the CT image while $pred$, is the generated image. 
    \item Mean Squared Error (MSE)
    \begin{equation}\label{eq:mse}
        MSE(ref, pred) = \frac{1}{N} \sum_{i=1}^{N} |ref(i) - pred(i)|^{2}
    \end{equation}
    The mean-squared error largely penalizes deviations from the reference image due to the difference being squared.
    \item Normalized Mean Squared Error (NMSE)
    \begin{equation}
        NMSE(ref, pred) = \sqrt{\frac{{\sum_{i=1}^{N} |CT(i) - CBCT(i)|^{2}}}{\sum_{i=1}^{N} |CT(i)|^{2}}}
    \end{equation}
    The $NMSE$ gives the mean squared error while also factoring in the signal power. 
    \item Power-to-Signal Noise Ratio (PSNR)
    \begin{equation}
    \begin{split}
    PSNR(ref, pred) &= 20 \times log_{10}(ref_{max})  \\
    & \quad - 10 \times log_{10}MSE(ref, pred)
    \end{split}
    \end{equation}

    $ref_{max}$ refers to the maximum value of the $ref$ image. $MSE(ref, pred)$ is computed as described in Equation \ref{eq:mse}
    \item Structural Similarity
    Index Metric (SSIM) \\
    $SSIM(ref, pred)$ is computed using the formula presented below, with $ref$ denoted as $x$, and $pred$ as $y$.
    \begin{equation}
        SSIM(x, y) = \frac{(2\mu_{x}\mu_{y} + c_1)(2\sigma_{xy} + c_2)}{(\mu_x^2 + \mu_y^2 + c_1)(\sigma_x^2 + \sigma_y^2 + c_2)}
    \end{equation}
    where $\mu$ and $\sigma$ represent the mean and variance respectively. $c_1$ and $c_2$ are variables used to stabilize division.
\end{itemize}
\subsection{Qualitative Inspection}\label{sec:methods:qualitative}
Qualitative criteria are  incorporated into the evaluation procedure, mainly due to the limitations of existing quantitative criteria in capturing \textit{undesired} effects of unpaired translation. Effects such as checkerboard patterns, addition of artifacts, modification of anatomies are not directly captured by metrics. For example, consider a model that translates images very well and offers good quantitative scores across all metrics. However, this model adds small artifacts such as air pockets that were not present in the original image. As a result, no matter how good the metric score, this model will not be accepted clinically. Therefore, qualitative evaluation and analysis, is inescapable. 

Structured qualitative inspection can allow comparing models in a more systematic manner. Through analyzing translations from various experiments, a set of criteria for qualitative inspection are formulated,
\begin{enumerate}
    \item \textit{Closeness to the target domain}:
    We attempt to determines how visually similar the translation looks to the domain of the target. For CBCT-CT, we do this by inspecting the images using soft-tissue windows. A soft-tissue window filters HU values outside the range of $(-135, 215)$ and allows better visual examination of soft-tissue boundaries. As a result, within these windows, it is possible to determine if different regions such as the heart, skeletal muscle, fat, etc, correspond visually between the translated image and target. We inspect specifically, (1) bony regions, (2) heart and lung, and (3) skeletal muscles. In addition to visual correspondence, we also check the reduction of CBCT artefacts in the translated image by comparing it with a region in the CBCT where artefacts were largely present. For \textit{maps $\leftrightarrow$ aerial photo}, we look at how well the translated image reproduces water bodies, highway roads, among other structures present in the maps image. 
    \item \textit{Presence of artifacts or undesirable elements}:
    Induction of artifacts are an established drawback of generative modelling with GAN-based methods \citep{zhang2019detecting}. Such artifacts are hard to identify using pixel-based quantitative metrics and no other metric that fully captures the range of possible artifacts in a CycleGAN is available yet. To this end, we inspect images manually to check for artifacts or any undesirable elements such as localized checkerboard artifacts that appear randomly.
    \item \textit{Quality of image in terms of resolution}:
    This criteria aims to identify reduction in perceived image quality for translated images. Some very commonly seen phenomena in CycleGAN translations are blurry images, aliasing-like effects and bright spots in parts of the image. Ideally, a reader study would be performed to analyze these factors. However, that is beyond the scope of this thesis.
\end{enumerate}

All the criteria mentioned above are inspected using mid-axial, sagittal and coronal views. Visualizing the entire 3D scan for 18 patients in the test set for multiple trained models would be very exhausting and impractical. Looking at three cross-sectional planes allows us to make a good judgement on the overall quality of the image. 

\subsection{Out-of-distribution Analysis}\label{sec:methods:ood}
Phantoms, as described in Section \ref{sec:methods:phantoms}, are used to test models on out-of-distribution data. Image-similarity metrics are computed on the phantom with an available body mask to generate quantitative metrics. Qualitative inspection of the phantom is done similar to the patient data, as mentioned in Section \ref{sec:methods:qualitative}. Special attention is paid to the translation of the tumour in the phantom as it is hypothesized to be a potential source of failure for the translation.

\subsection{Domain-specific Evaluation}\label{sec:methods:domain_eval}
Apart from the quantitative and qualitative evaluations highlighted above, we would like to understand if the methods proposed as a part of the thesis benefit the use-case of adaptive radiotherapy, for which we chose to design such methods. In order to establish this, we do a short analysis of HU value distributions between the original, target and the improved CBCT (translated scan). We also compare these scans based on line profiles that demonstrate HU values observed when a line passing through the heart, lung, skeletal muscles and bones, is drawn in the axial plane.   \par
Additionally, the improved CBCT can also be used to generate RT contours through incorporation of automated segmentation methods. To check if the CBCT improvement benefits this task, we compare the difference in segmentation contours obtained between original and generated images. An automated lung segmentation model\footnote{https://github.com/JoHof/lungmask} \citep{Hofmanninger2020}, trained on a large and diverse dataset, is used to segment the CT/CBCT scan into left and right lung. Since the test data contains ground truth contours, we generate automated contours for the CT, original CBCT and improved CBCT and compare each with the ground truth using the Dice score. 

\subsection{Validation and Selection of Best Checkpoints}
\begin{figure}[H]
    \centering
    \includegraphics[width=0.5\textwidth]{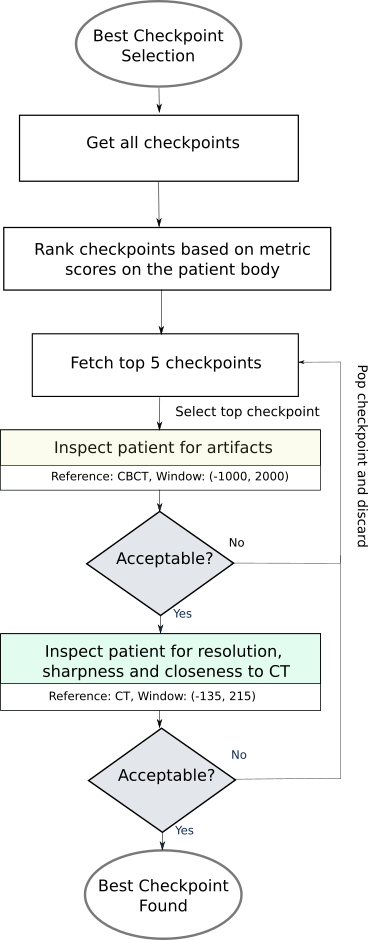}
    \caption{Process of selection of best checkpoint using the validation data. For each of the inspection boxes, a reference and window is outlined that represents what it was compared against and the viewing window used for comparison}
    \label{fig:methods:checkpoints}
\end{figure}

Unpaired GAN-based training can offering varying translations across different iterations. To pick the best performing model (quantitatively and qualitatively), a well designed validation method is needed. In this thesis, a validation method that analyzes translation at different checkpoints quantitatively, followed by visual inspection is employed. Checkpoints are ranked in terms of their metric scores on the patients in the validation set. The top-5 ranked checkpoints are taken as candidates. From this list of candidates, the top checkpoint is popped and if it satisfies visual inspection, is chosen as the best checkpoint. If it fails the inspection, it is discarded and the next checkpoint is considered. For visual inspection, criteria that are outlined in Section \ref{sec:methods:qualitative} are used. Figure \ref{fig:methods:checkpoints} shows a flowchart diagram describing this process of checkpoint selection. 

\chapter{Experiments and Results}
In this section, experimental configurations that are used to investigate the research questions posed as a part of this thesis are outlined. Each experiment is performed as an ablation study with only the parameter that is under-investigation being changed. However, in some cases, when this is not possible, for example, due to hardware and other limitations, its implications on the experiment are discussed. 

\section{Experimental Setup}
Firstly, two base configurations are determined, 
\begin{enumerate}
    \item For models trained with the \textit{maps $\leftrightarrow$ aerial photo} data, called the \textit{2D base configuration}
    \item For models trained with the CBCT-CT data, called the \textit{3D base configuration}
\end{enumerate}
 
Table \ref{tab:expts:configuration} shows the parameters for the two base configurations in detail. 
\begin{table}
\centering
\begin{tabular}{lll}
\hline
                                                                               & 2D Configuration                 & 3D Configuration               \\ \hline \vspace{2mm}

Generator                                                                      & UNet 2D                          & VNet 3D                        \\ \vspace{2mm}

Discriminator                                                                  & PatchGAN 2D                      & PatchGAN 3D                    \\ \vspace{2mm}

LR                                                                             & D: 0.0001, G:0.0002              & D: 0.0002, G:0.0004            \\ \vspace{2mm}

Batch-size                                                                     & 4*                               & 1                              \\ \vspace{2mm}

LR schedule                                                                    & \multicolumn{2}{l}{Fixed for 50\%, Linear decay for 50\%}          \\ \vspace{2mm}

Optimizer                                                                      & \multicolumn{2}{l}{Adam $(\beta_{1} = 0.5,\ \beta_{2} = 0.999)$} \\ \vspace{2mm}

\begin{tabular}[c]{@{}l@{}}Lambda \\ ($\lambda_{A}, \lambda_{B}$)\end{tabular} & 10                               & 5 \\ \vspace{2mm}

Input size      & $(256, 256)$ & $(16, 320, 320)$
\\ \vspace{2mm}

Normalization &  \multicolumn{2}{l}{Instance normalization} \\ \vspace{2mm}

Training iterations & 219.2k & 30k \\ \hline \vspace{2mm}

\end{tabular} \\
\small{$^{*}$ Batch-size of 4 is used as initial experiments with higher batch-size did not show any improvement.} \\

\caption{Base configurations used for experiments performed across the two datasets}
\label{tab:expts:configuration}

\end{table}
The 2D models are trained on a GTX1080Ti with 4GB of memory for 219.2k iterations (200 * size of dataset) while the 3D models are trained on Tesla V100 with 16GB memory for 30k iterations. All experiments are run with mixed precision enabled, unless mentioned otherwise, in order to efficiently use available GPU memory.

\subsection{Network Architectures}
In this section, we describe in detail the network architectures used for 2D and 3D configurations. Specifically, the generator and discriminator architectures along with their parameters are discussed.
\subsubsection{2D Configuration}\label{sec:expts:2d_conf}
The \textit{2D base configuration} consists of a 2D UNet \citep{Ronnerberger2015} as generator and PatchGAN2D \citep{Isola2016} as discriminator. The 2D UNet consists of a down-sampling path with 7 convolutional blocks, each with convolutions of kernel size 4 and stride 2. Following the convolution, a LeakyReLU and instance norm is applied (except for the terminal levels where LeakyReLU is applied for the input level and ReLU is applied for the output level following the convolution). This is followed by an up-sampling path that consists of the same number of blocks as the down-sampling. Each of the blocks consist of transposed convolutions of the same kernel size and stride as the convolutions. After the transposed convolution, instance norm and ReLU layers follow (except for the last block where a $tanh$ follows the transposed convolution). Between each of the down-sampling and up-sampling blocks of matching feature sizes, a skip-connection is made which concatenates their feature maps. Figure \ref{fig:appendix:2d} in the Appendix shows a block diagram of the 2D UNet generator. The PatchGAN2D consists of 5 convolutional blocks with instance norm and LeakyReLU after each of the convolutions with kernel size 4 and stride 2 (except for the input block where no instance norm is applied and the output block where neither instance norm nor LeakyReLU is applied). For the given input size used for our 2D configuration, the PatchGAN2D outputs a patch of size $30\times30$. 

\subsubsection{3D Configuration}\label{sec:expts:3d_conf}
The \textit{3D base configuration} consists of a 3D VNet \citep{milletari2016vnet} as generator and PatchGAN3D as discriminator. The 3D VNet structure consists of an input block, 4 down-sampling blocks, 4 up-sampling blocks and an out block. The input block consists of a 3D convolutional block of kernel size 5 followed by instance norm and PReLU. The four down-sampling blocks consists of 1, 2, 3 and 2 convolutional blocks, respectively, with varying kernel sizes and strides. The four up-sampling blocks consists of 2, 2, 1, 1 convolutional blocks.  The output block contains 2 convolutional blocks, the first followed by an instance norm and PReLU and the second followed by a $tanh$.  This configuration is arrived at based on previous experiments conducted on other medical imaging data with promising performance. Skip connections similar to the UNet 2D are also seen in the 3D VNet. Figure \ref{fig:appendix:3d} in the Appendix shows a block diagram of the 3D VNet generator. The PatchGAN 3D is a 3D version of the PatchGAN 2D, obtained by replacing the the 2D convolutions with 3D convolutions.

\section{Exploring Default CycleGAN Constraints}\label{sec:expts:RQ1}
In this experiment, the significance of $\lambda$ parameter, that determines the balance between cycle-consistency and adversarial losses, is investigated. The authors of CycleGAN show that excluding the cycle-consistency distorts the translated images with hallucinated structures and patterns. However, they do not discuss the impact of different values of lambda on translation. We conduct experiments with different values of lambda, specifically $\lambda=1$ and $\lambda=50$, with the end goal of  gauging their impact on translation performance. These values are chosen as extremes as the default value of $\lambda=10$ is chosen by the authors. As we expect to gain an intuition of $\lambda$ values and their impact on translation in a general sense, we run the experiments only on \textit{maps $\leftrightarrow$ aerial photo} dataset with the 2D configuration. 
\subsection{Results}
\begin{figure}[H]
    \centering
    \includegraphics[width=0.8\textwidth]{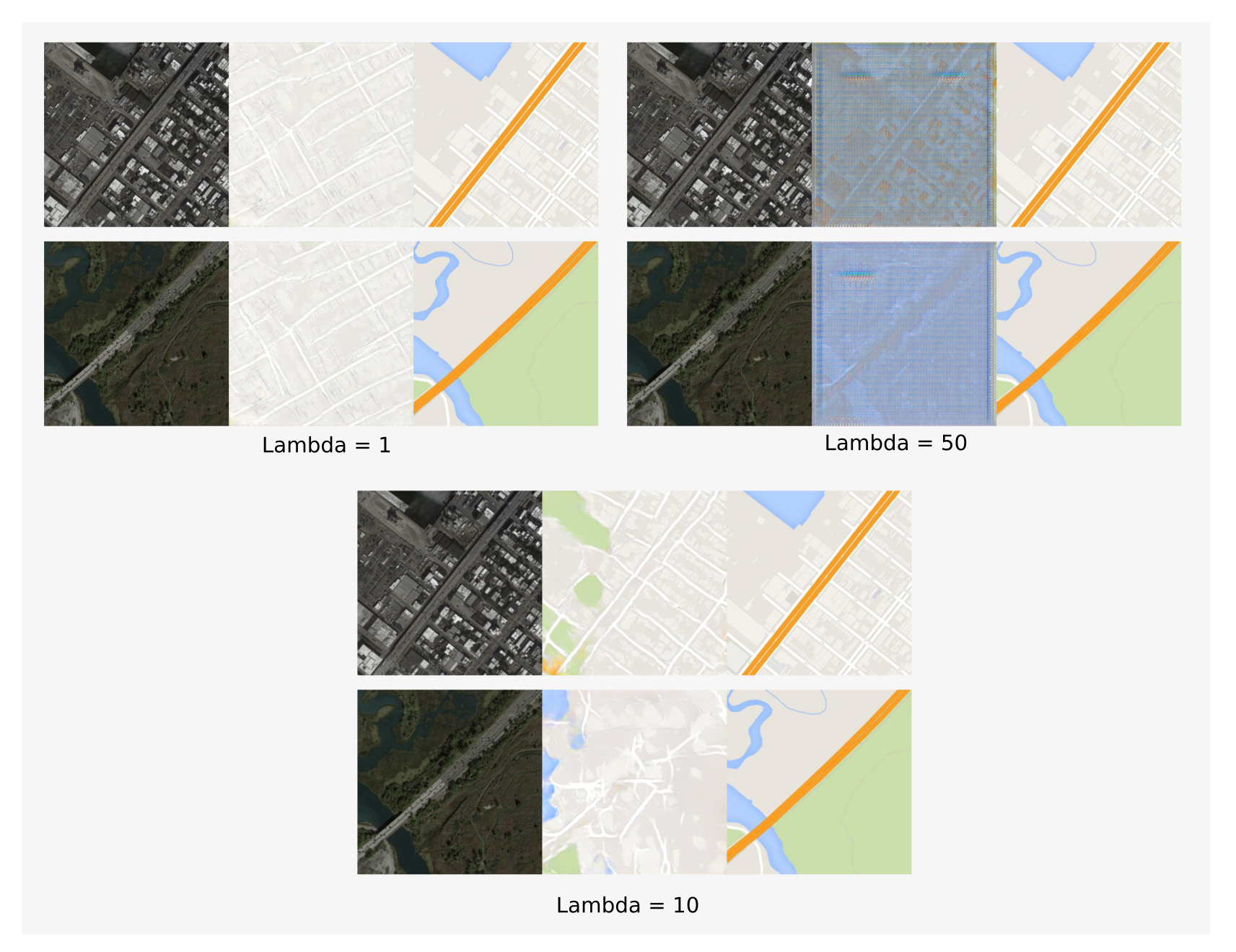}
    \caption{Translated images for different values of $\lambda$. For each parameter, two sets of images are shown. An image set consists of the source image in the first column, translated image in the second and ground truth in the third.}
    \label{fig:results:exp1}
\end{figure}

Table \ref{tab:results:exp1} and Figure \ref{fig:results:exp1} show the quantitative and qualitative results of different $\lambda$ parameters. 
\begin{table}[H]
\centering
\begin{tabular}{llllll}
\hline
\textbf{Model} & \textbf{MAE}   & \textbf{MSE}   & \textbf{NMSE}  & \textbf{PSNR}   & \textbf{SSIM}  \\ \hline
$\lambda=1$    & 0.113          & 0.030          & 0.054          & \textbf{17.036} & 0.213          \\ 
$\lambda=10$   & \textbf{0.111} & \textbf{0.030} & \textbf{0.053} & 16.549          & \textbf{0.242} \\ 
$\lambda=50$   & 0.563          & 0.363          & 0.551          & 4.656           & 0.043 \\ \hline
\end{tabular}
\caption{Image similarity metrics on the test set for experiments with different values for the $\lambda$ parameter on the \textit{maps $\leftrightarrow$ aerial photo} dataset}
\label{tab:results:exp1}
\end{table}

$\lambda=10$ performs the best quantitatively, providing best scores for all metrics except PSNR. $\lambda=1$ provides the best PSNR score. Looking at Figure \ref{fig:results:exp1}, we see that $\lambda=10$ also performs the best qualitatively by estimating several components of the image satisfactorily. However, it is far from being geometrically accurate to the structures in the target image. Translations from $\lambda=1$ undergo mode collapse with different input images being mapped to the same output translation. Mode collapse is a common occurence in GANs and avoiding it is an active area of research. $\lambda=50$ shows structures that correspond best with the target image, but do not identify as being the maps domain. A lot of high-frequency components are seen in the translated image. \\

\begin{figure}[H]
    \centering
    \includegraphics[width=\textwidth]{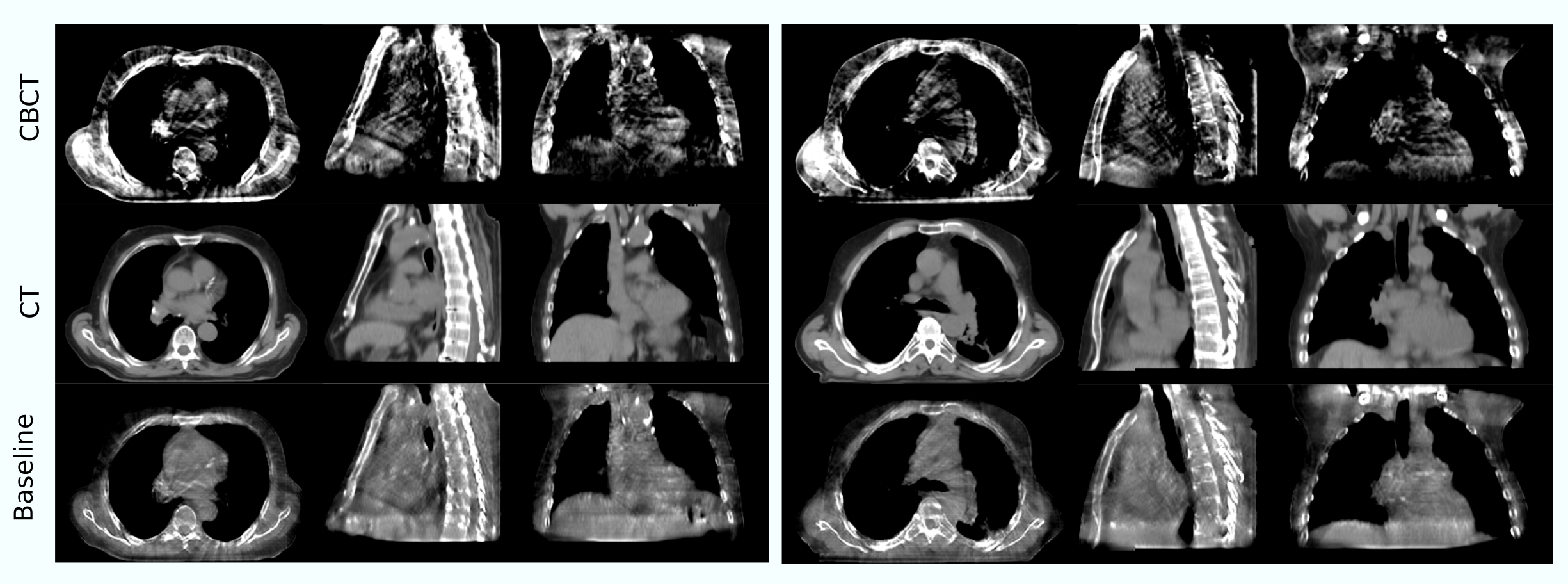}
    \caption{Mid-axial, sagittal and coronal views for CBCT, CT and baseline model on two different patients from the test, shown using the soft-tissue window. The two patients are separated by a white space.}
    \label{fig:results:baseline}
\end{figure}

\textbf{CBCT-CT Dataset}
The baseline model, with $\lambda=10$, is run on the CBCT-to-CT dataset and the quantitative results can be seen in Table \ref{fig:results:exp2} while the translated images for 2 patients, chosen randomly, can be seen in Figure \ref{fig:results:baseline}. Only soft-tissue windows are shown as they are easier to inspect for quality of translation. However, it is harder to inspect addition of GAN-artifacts or anatomy modifications in this window range. Since no major artifacts were observed on the baseline, we present only the windowed images.  

The baseline offers satisfactory translation and looks closer to the CT domain than the CBCT domain. It retains the structure of the original scan, unlike what was seen on the aerial imagery. However, zooming in on the images shows that there are checkerboard artifacts, especially visible in the heart region on the axial view (left). We also see that CBCT scattering artefacts are still retained in the translated images. This can be seen properly in the sagittal views (center). 

\section{Adapting Constraints For Improved Translation}\label{sec:expts:RQ2}
This experiment is designed in order to address the possibility of dynamically adapting $\lambda$ values based on training dynamics. The CycleGAN using adaptive $\lambda$ strategies is called the AdaptiveCycleGAN, for ease of reference. As mentioned in Section \ref{sec:methods:adaptive_control}, it is possible to use different \textit{control signals} to evaluate the state of training in a CycleGAN. A set of experiments are designed to investigate if these two control signals can be used for adaptive training. In these experiments, for (1) discriminator-based control signal, different change rates and $D_{train}$ vs $D_{generator}$ is experimented with and for (2) cycle-consistency loss, change modes and rates are varied. The configurations for the best performing models (chosen based on the image-similarity metric scores on the validation set) from these initial experiments can be found in Table \ref{sec:expts:configs}.\\

\begin{table}
\begin{tabular}{lll}
\hline
                    & \makecell{AdaptiveCycleGAN \\ discriminator-based} & \makecell{AdaptiveCycleGAN \\ cycle-consistency based} \\ \hline
Change rate         & 0.01                & 0.01                    \\
Change mode         & -                   & Threshold-based         \\
Frequency of update & 2000                & 1000                    \\
Sample Size         & 256                 & -                       \\
$r_t$               & 0.6                 & -                       \\
Patience            & -                   & 100                     \\
Warmup iterations   & -                   & 5000                    \\ \hline
\end{tabular}
\caption{Best performing configurations determined across different AdaptiveCycleGAN experiments.}
\label{sec:expts:configs}
\end{table}

The full range of experiments can be found as a Weights and Biases workspace \footnote{https://wandb.ai/surajpai/aerial\_to\_maps}. The best performing run across the two control signals are then run on the CBCT-CT dataset. 

\subsection{Results}
\begin{figure}[H]
    \centering
    \includegraphics[width=0.8\textwidth]{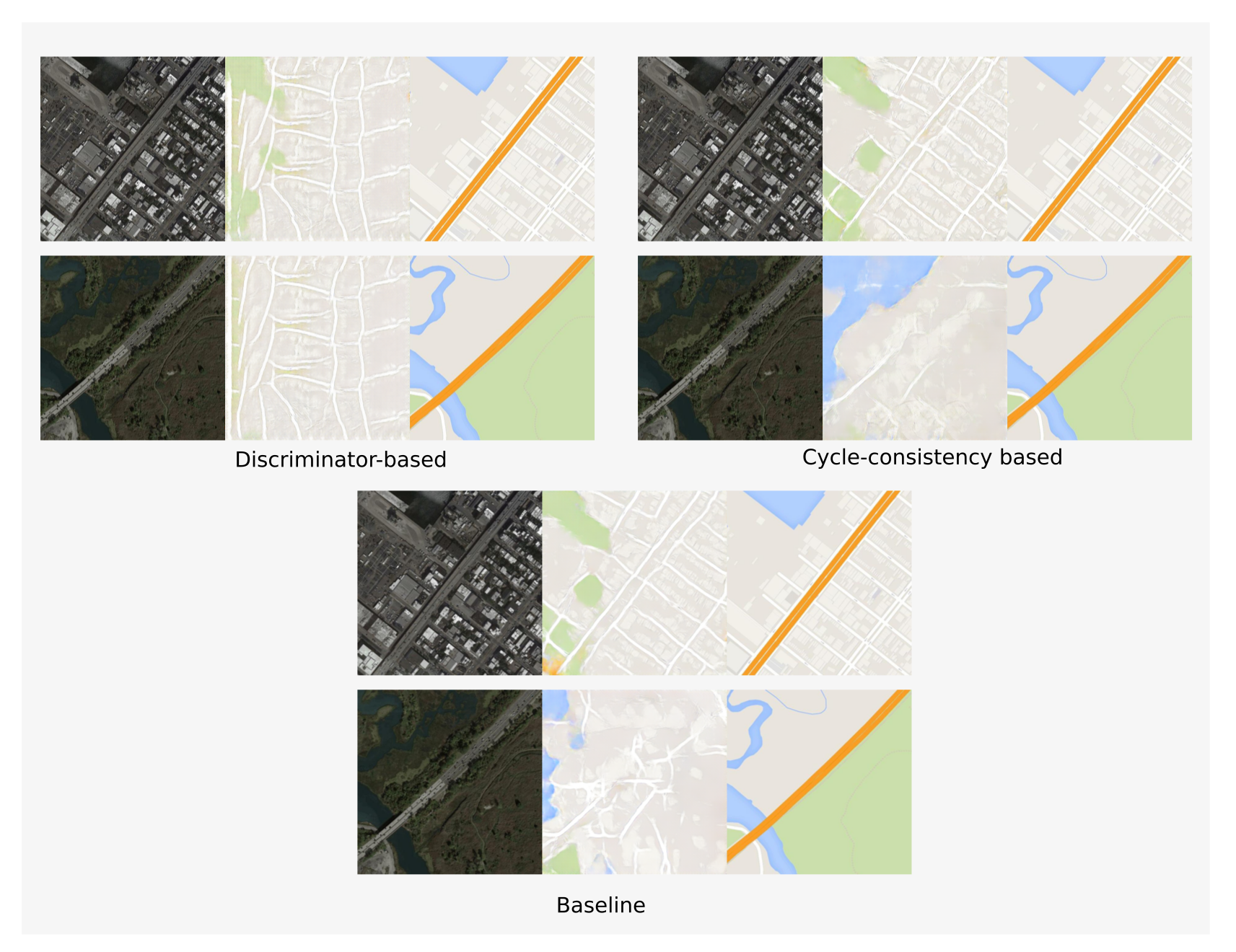}
    \caption{Source, translated and target images for baseline, discriminator-based and cycle-consistency based adaptive $\lambda$ strategies. For each experiment, two sets of images are shown. An image set consists of the source image in the first column, translated image in the second and ground truth in the third.}
    \label{fig:results:exp2}
\end{figure}

Table \ref{tab:results:exp2} and Figure \ref{fig:results:exp2} shows the quantitative and qualitative results of the two control signals investigated for adaptive $\lambda$ strategies on the \textit{maps $\leftrightarrow$ aerial photo} dataset.

\begin{table}[H]
\centering
\begin{tabular}{llllll}
\hline
\textbf{Model}          & \textbf{MAE}   & \textbf{MSE}   & \textbf{NMSE}  & \textbf{PSNR}   & \textbf{SSIM}  \\ \hline
Baseline                & 0.111          & \textbf{0.030} & 0.053          & \textbf{16.549} & \textbf{0.242} \\ \vspace{2mm} 
\makecell[l]{AdaptiveCycleGAN \\ discriminator-based}     & 0.122          & 0.034          & 0.058          & 15.759          & 0.197          \\ \vspace{2mm}
\makecell[l]{AdaptiveCycleGAN \\ cycle-consistency based }& \textbf{0.110} & \textbf{0.030} & \textbf{0.052} & 16.531          & 0.239          \\ \hline
\end{tabular}
\caption{Image similarity metrics on the test set for experiments to determine the best control signal for adaptive $\lambda$ strategies on the \textit{maps $\leftrightarrow$ aerial photo} dataset.}
\label{tab:results:exp2}
\end{table}

The baseline model achieves the best PSNR and SSIM, while the cycle-consistency based AdaptiveCycleGAN obtains the best MAE and NMSE. Both of them obtain very similar MSE values. The discriminator-based AdaptiveCycleGAN performs much worse compared to baseline and cycle-consistency control signal. 
Qualitative analysis of the translated images shows very comparable properties between baseline and cycle-consistency based AdaptiveCycleGAN. Specific attributes that maybe favoured from either model exist in the translated images. For instance, the baseline model, provides less blurrier additions (blur added in 2nd image set for the cycle-consistency AdaptiveCycleGAN) to the image while the cycle-consistency AdaptiveCycleGAN adds lesser hallucinated structures (as seen in 2nd image set for the baseline). The discriminator-based AdaptiveCycleGAN has mode collapsed with meaningless representations. \\

\textbf{CBCT-CT Dataset}
The cycle-consistency based AdaptiveCycleGAN is tested on the CBCT-CT dataset, and for simplicity, it is addressed as AdaptiveCycleGAN henceforth. Table \ref{tab:results:exp3} and Figure \ref{fig:results:exp4} show the quantitative and qualitative results respectively.
\begin{table}[H]
\begin{tabular}{llllll}
\hline
\textbf{Model}   & \textbf{MAE}    & \textbf{MSE}       & \textbf{NMSE}    & \textbf{PSNR}   & \textbf{SSIM}   \\ \hline
Baseline         & \textbf{88.846} & \textbf{24244.151} & \textbf{0.03086} & \textbf{29.365} & \textbf{0.9349} \\
AdaptiveCycleGAN & 101.94          & 24899.015          & 0.03171          & 29.193          & 0.9276  \\ \hline   
\end{tabular}
\caption{Image similarity metrics on the test set for the AdaptiveCycleGAN experiment run on the CBCT-CT dataset. Note that all values with units are in a CT number scale (between 0 to 3000)}
\label{tab:results:exp3}
\end{table}

\begin{figure}[H]
    \centering
    \includegraphics[width=\textwidth]{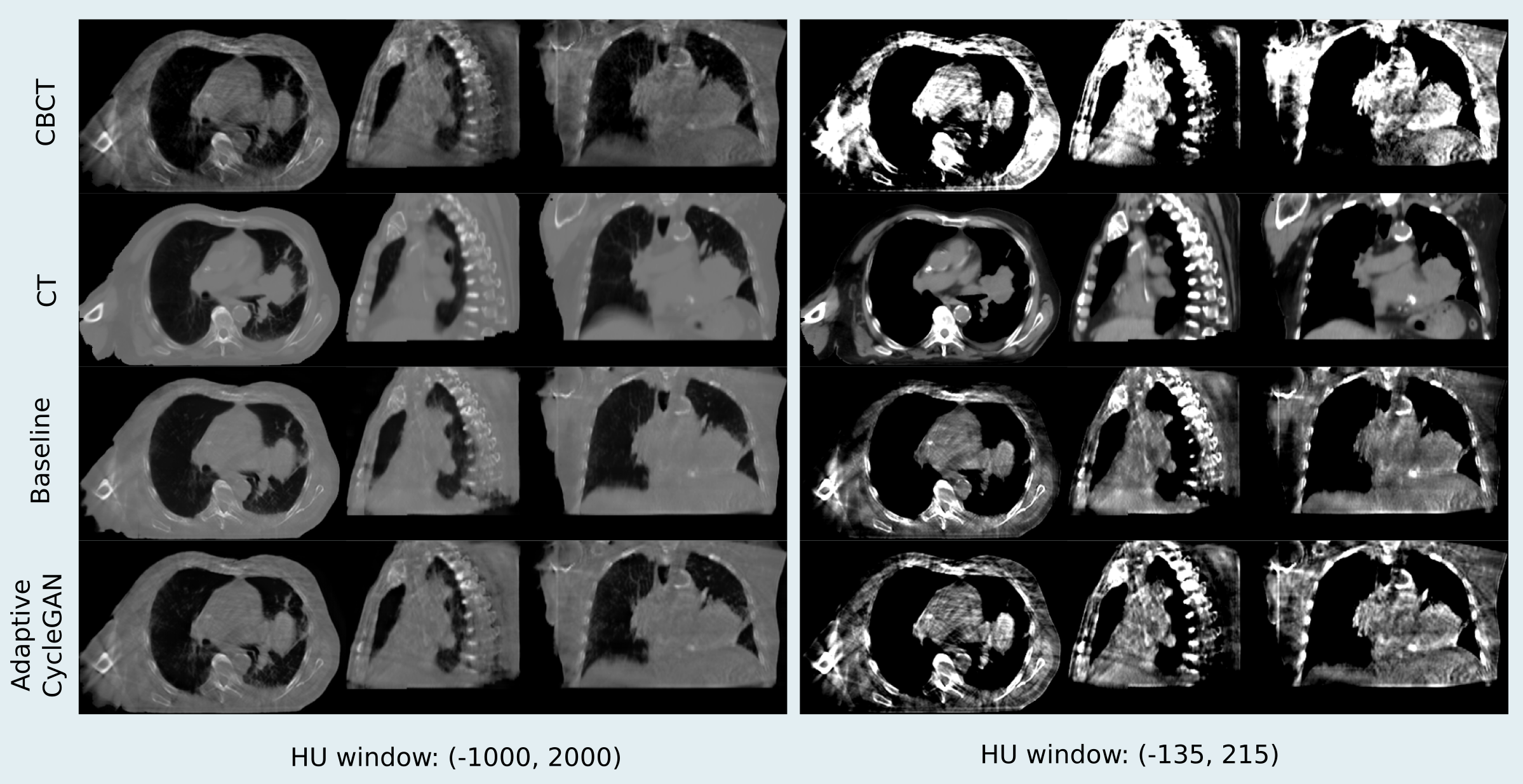}
    \caption{Mid-axial, sagittal and coronal views for CBCT, CT , baseline and AdaptiveCycleGAN generated images for a patient chosen randomly from the test set. Two different HU windows are used to since the soft-tissue window (-135, 215) does not sufficiently show addition of artifacts.}
    \label{fig:results:exp4}
\end{figure}
The AdaptiveCycleGAN obtains better MSE and NMSE while the baseline obtains better MAE, PSNR and SSIM. Do note that baseline trumps the AdaptiveCycleGAN in MAE and SSIM by a considerable margin. This can also be seen visually in Figure \ref{fig:results:exp4} where the AdaptiveCycleGAN is closer to the CBCT (source) rather than the CT (target). However, a lot of noise and CBCT artefacts are still present. No GAN-induced artifacts are observed (a checkpoint @ 14k iterations was chosen as later checkpoints generated plenty of artifacts). In contrast, the baseline gets much closer to the CT distribution and suppresses CBCT artefact in many regions across, especially in the heart and skeletal muscles. 

\section{Data-driven Constraints for CBCT Image Quality Enhancement}\label{sec:expts:RQ3}
After seeking to adapt existing constraints, the next set of experiments look at adding more constraints to the CycleGAN framework. This is with the hypothesis that existing constraints still have issues with maintaining structure between the two domains. Section \ref{sec:methods:structure_losses} describes the different data-driven constraints, in the form of structure losses, investigated as a part of the thesis.  \par
For the structure losses, experimental configurations that differ from the base configuration or that may be of specific interest are mentioned below,  \par
\textbf{MIND loss:} A few changes were made in the experiment configuration for the MIND loss, (1) the authors propose a weight of $\lambda_{structure}=5$, this is changed to $\lambda_{structure}=50$ through scale-matching with other losses, (2) a patch-size of $(16, 192, 192)$ is used for the MIND loss due to memory restrictions, as a direct consequence of (3) Mixed-precision being unsuitable for this experiment as it led to early divergence. \\
\textbf{Generalized Frequency loss:} Two different distance metrics were used for generalized frequency loss, shown in Equation \ref{eq:methods:gen_freq_loss}, the $L_1$ distance between the frequency representations and the $L_2$ distance between the frequency representations. Other distance metrics such as $L_p$ distances may also offer interesting properties but are not experimented with in this thesis due to time and resource constraints. \\
\textbf{Registration loss:} As the values of the registration loss measure was not bound, different methods of normalization of the losses were tried, namely, (1) $tanh$ squashing, (2) loss-clipping and (3) full range (not normalized). \\
\textbf{Combined Loss:} A combination loss consisting of the Frequency L1 loss with the MIND loss is investigated. The losses have $\lambda_{struct}$ values consistent with their individual experiments, and are summed to obtain the combined loss. This is trained with a patch size of $(16, 320, 320)$. 

A total of six different models are compared in this experiment, (1) Baseline, (2) MIND, (3) Frequency L1, (4) Frequency L2, (5) MIND + Frequency L1 and (6) Registration loss

\subsection{Results}
Table \ref{tab:results:exp4} and Figure \ref{fig:results:exp4} present the different image similarity metrics and visuals of a patient scan from different experiments. Note that the registration loss experiment is not shown due to its complete failure to converge to a meaningful translation. Visuals of its failure modes are presented in Figure \ref{fig:results:reg_fail} and subsequently discussed. 
\begin{table}[H]
\begin{tabular}{llllll}
\hline
\textbf{Model}        & \textbf{MAE}    & \textbf{MSE}       & \textbf{NMSE}    & \textbf{PSNR}   & \textbf{SSIM}   \\ \hline
Baseline              & 88.846          & 24244.151          & 0.03086          & 29.365          & 0.9349          \\
MIND                  & 85.911          & 25604.077          & 0.03249          & 29.265          & \textbf{0.9441} \\
Frequency loss $L_1$  & \textbf{85.503} & \textit{20433.408} & \textit{0.02606} & \textit{30.016} & 0.935           \\
Frequency loss $L_2$  & \textit{85.966} & \textbf{20247.973} & \textbf{0.02577} & \textbf{30.121} & \textit{0.9387} \\
MIND + Frequency loss & 86.632          & 21125.905          & 0.02696          & 29.88           & 0.9346          \\ \hline
\end{tabular}
\caption{Quantitative metrics obtained on the test set for experiments with various structure losses run on the CBCT-CT dataset. Note that all values with units are in a CT number scale (between 0 to 3000)}
\label{tab:results:exp4}
\end{table}

\begin{figure}[H]
    \centering
    \includegraphics[width=\textwidth]{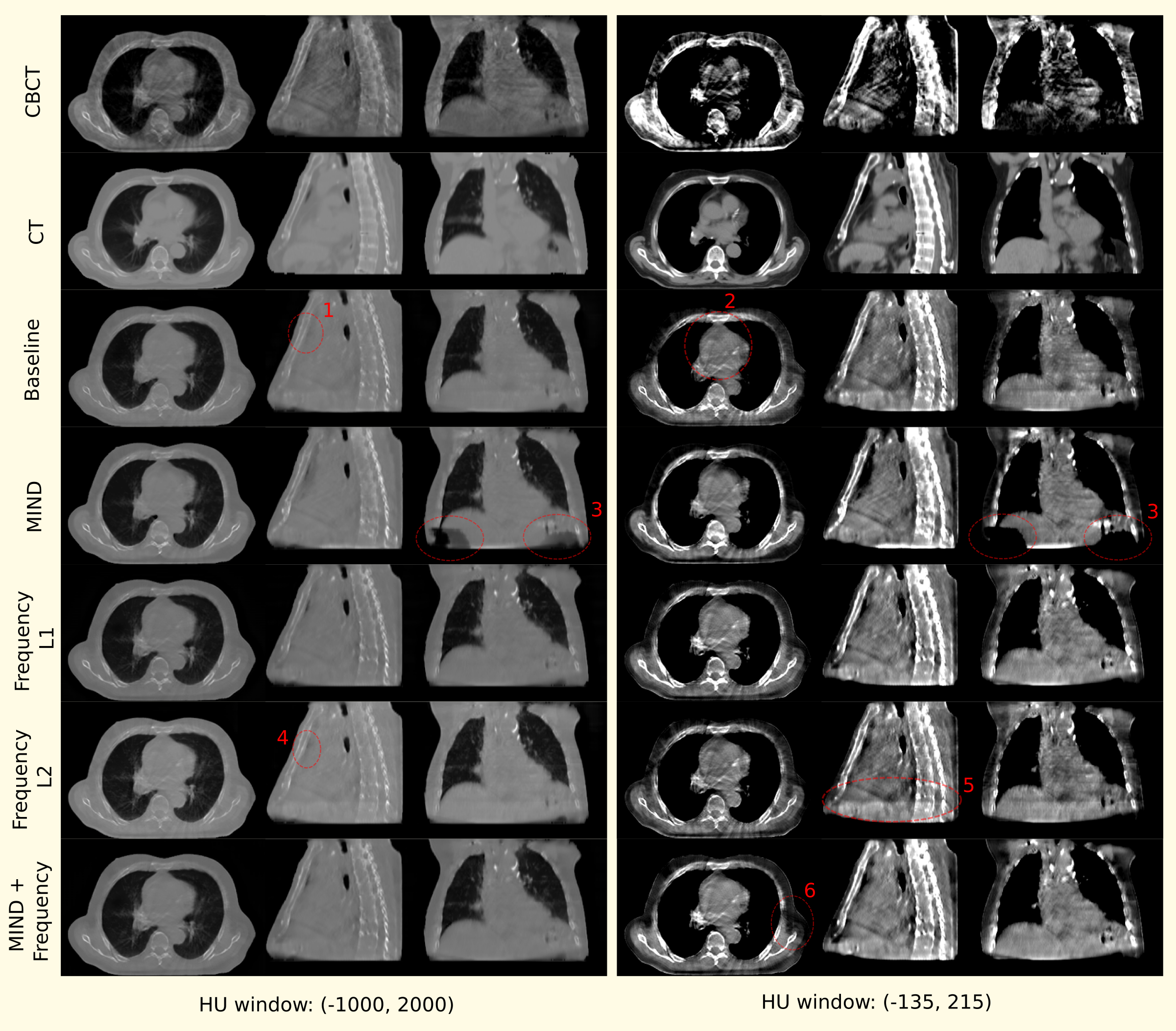}
    \caption{Mid-axial, sagittal and coronal views for CBCT, CT, and generated images from models with different data-driven constraints for a patient chosen randomly from the test set. Two different HU windows are used to since the soft-tissue window (-135, 215) does not sufficiently show addition of artifacts. Qualitative observations discussed in-text are marked with red dotted ellipses and numbered. }
    \label{fig:results:exp4}
\end{figure}
The baseline model is one of the worst performing models with the worst MAE and second worst SSIM, MSE, NMSE and PSNR. MIND loss provides the best SSIM but gives the worst MSE, NMSE and PSNR among all the models investigated. In-terms of MSE, NMSE and PSNR, the Frequency loss with $L_2$ performs the best.
Inspecting the scans visually, a sample of which is shown in Figure \ref{fig:results:exp4}, we make the following observations based on criteria highlighted in Section \ref{sec:methods:qualitative}, \footnote{The observations are also pointed out in the Figure \ref{fig:results:exp4} using red dotted ellipses and numbering. The numbering scheme in the list corresponds with numbers in the image} 
\begin{enumerate}
    \item Air pockets that are present in the original scan are closed by the baseline model. 
    \item For the baseline model, decrease in the quality of the translated image is observed through addition of checkerboard-like patterns.
    \item MIND loss adds unexplained artifacts in the form of black density reduction fields. 
    \item Frequency $L_2$ also closes air pockets similar to the baseline model.
    \item Frequency $L_2$ provides a shift in density as we move down to the diaphragm, as seen on the sagittal view. 
    \item MIND + Frequency $L_1$ causes a random drop in density across a particular region. 
\end{enumerate}
The above observations are made across multiple patients from the test dataset. For brevity, only features from a single patient are described. Although the MIND loss seems to visually be closest to the CT, it adds significant artifacts. The next closest candidate where no artifacts or image quality drops are observed, is the Frequency $L_1$.

\subsubsection{Failure of Registration Loss}
\begin{figure}[H]
    \centering
    \includegraphics[width=0.8\textwidth]{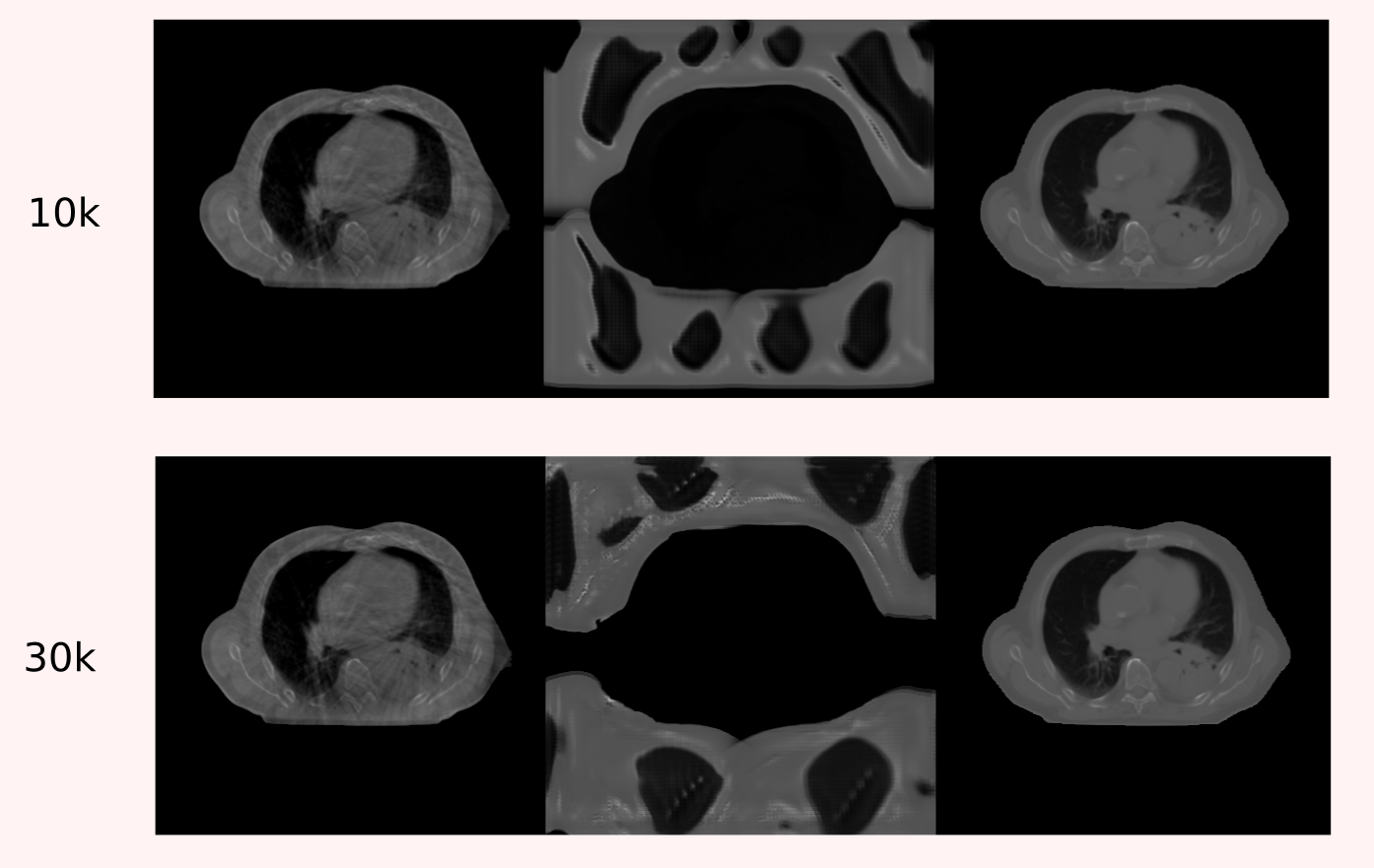}
    \caption{Failure modes of the registration loss shown on a patient from the validation set at different iterations during training. The columns show source, generated and target images respectively.}
    \label{fig:results:reg_fail}
\end{figure}

Multiple configurations of registration loss, as mentioned in Section \ref{sec:expts:RQ3}, were tested along with hyper-parameter searches. None of them seemed to provide any meaningful translations. Figure \ref{fig:results:reg_fail} shows the registration loss at two checkpoints during training.
From Figure \ref{fig:results:reg_fail}, it is evident that the training is divergent at 10k iterations and gets worse with adding `bony` structures toward the end at 30k iterations.  Note that all other configurations performed similar to the figure, if not, worse. The registration loss was not able to be utilized directly in the CycleGAN framework and also failed with simple modifications. Further analysis on how this loss may be incorporated into the framework without divergence needs to be conducted. 

\subsubsection{Out-of-distribution Analysis}
\begin{figure}[H]
    \centering
    \includegraphics[width=\textwidth]{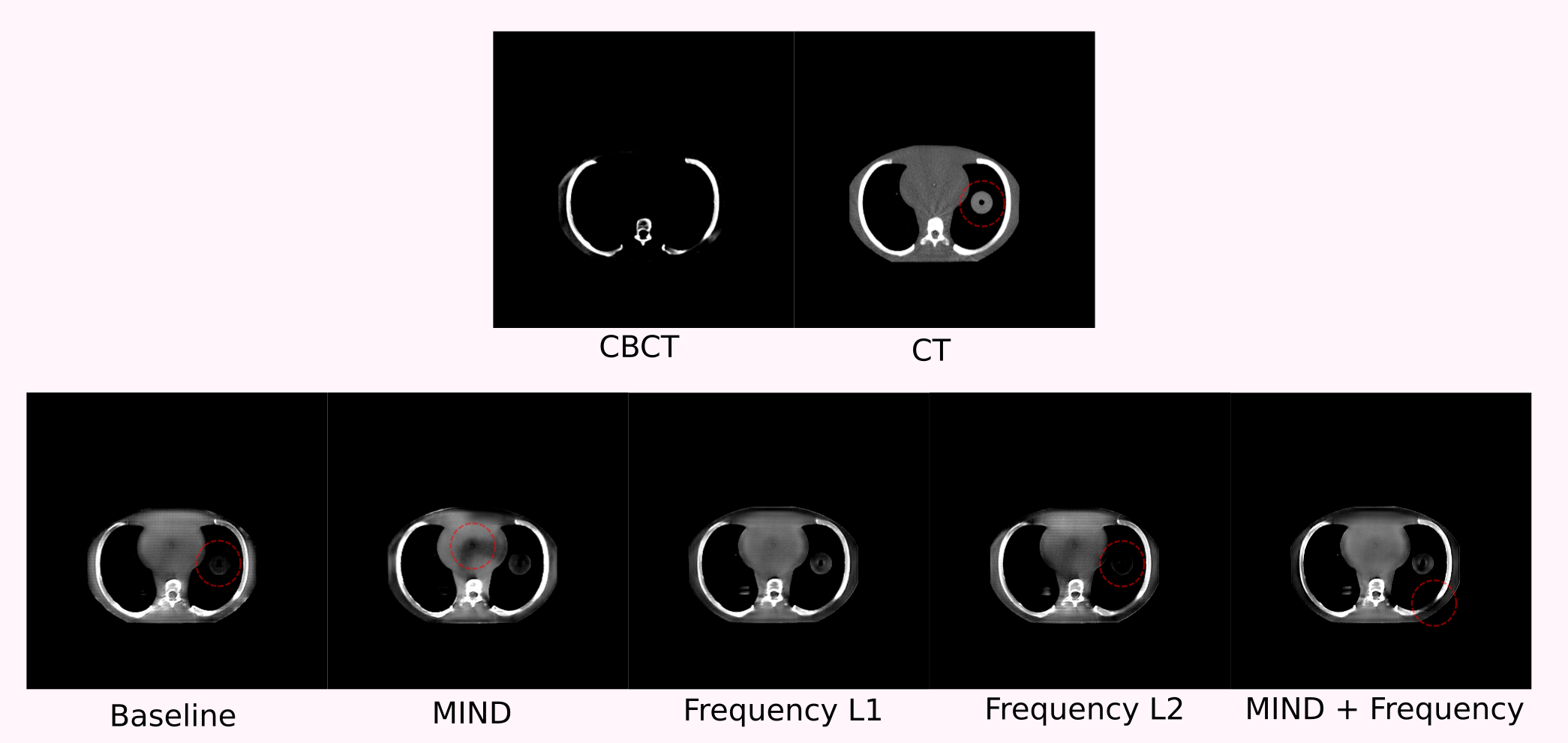}
    \caption{Mid-axial views of the CBCT and CT scans of the phantom shown along with generated images from models with different structure losses. Red dotted circles are used to highlight qualitative observations that are discussed in-text.}
    \label{fig:results:phantom_ood}
\end{figure}

Table \ref{tab:results:phantom_ood} and Figure \ref{fig:results:phantom_ood} show metric scores and visuals of phantoms as outlined in Section \ref{sec:methods:ood}.
\begin{table}[H]
\begin{tabular}{llllll}
\hline
\textbf{Model}              & \textbf{MAE}    & \textbf{MSE}       & \textbf{NMSE}    & \textbf{PSNR}   & \textbf{SSIM} \\ \hline
Baseline                    & 72.161          & 16207.693          & 0.0243           & 34.551          & 0.9761        \\
MIND                        & \textbf{62.743} & \textbf{11303.564} & \textbf{0.01694} & \textbf{36.116} & 0.9852        \\
Frequency loss $L_1$        & 71.387          & 16878.252          & 0.0253           & 34.375          & 0.9757        \\
Frequency loss $L_2$        & 63.651          & 12046.446          & 0.01806          & 35.84           & 0.9826        \\
MIND + Frequency loss $L_1$ & 75.337          & 17723.257          & 0.02657          & 34.163          & 0.9747        \\ \hline
\end{tabular}
\caption{Image similarity metrics on the phantom for experiments with various structure losses. Note that all values with units are in a CT number scale (between 0 to 3000)}
\label{tab:results:phantom_ood}
\end{table}
MIND loss shows the best scores across all metrics. However, when looking at the visuals for the phantom mid-axial slice, MIND loss does not correct values very accurately as can be seen with the black region in the center of the simulated heart. Frequency $L_2$ and baseline models do not correct tumour values well (as indicated by the red circles in the CT, Baseline and Frequency $L_2$ generated samples). Moreover, they also add checkerboard patterns, which can be observed by zooming in on the image. The MIND + Frequency shows regions that are much darker than the CT, as shown by the red circle. Frequency $L_1$, similar to observations on the patient data, provides robust translations on the phantom with neither GAN-induced artifacts nor loss of quality. It is interesting to note that the Frequency $L_1$ metric values are among the worst across all models. 

\section{Domain-specific Evaluation}\label{sec:expts:RQ4}
In this section we present the results of domain-specific evaluation criteria as described in Section \ref{sec:methods:domain_eval}. The Frequency loss with $L_1$, which is chosen as the best-performing model, due to its robust performance across both patient and phantom data, is used for subsequent evaluation. The translated scan generated using this model is termed as the sCT (Synthetic CT) and will be used to refer to it henceforth. 

\subsection{Histogram and Line Profiles}
\begin{figure}
    \centering
    \includegraphics[width=0.8\textwidth]{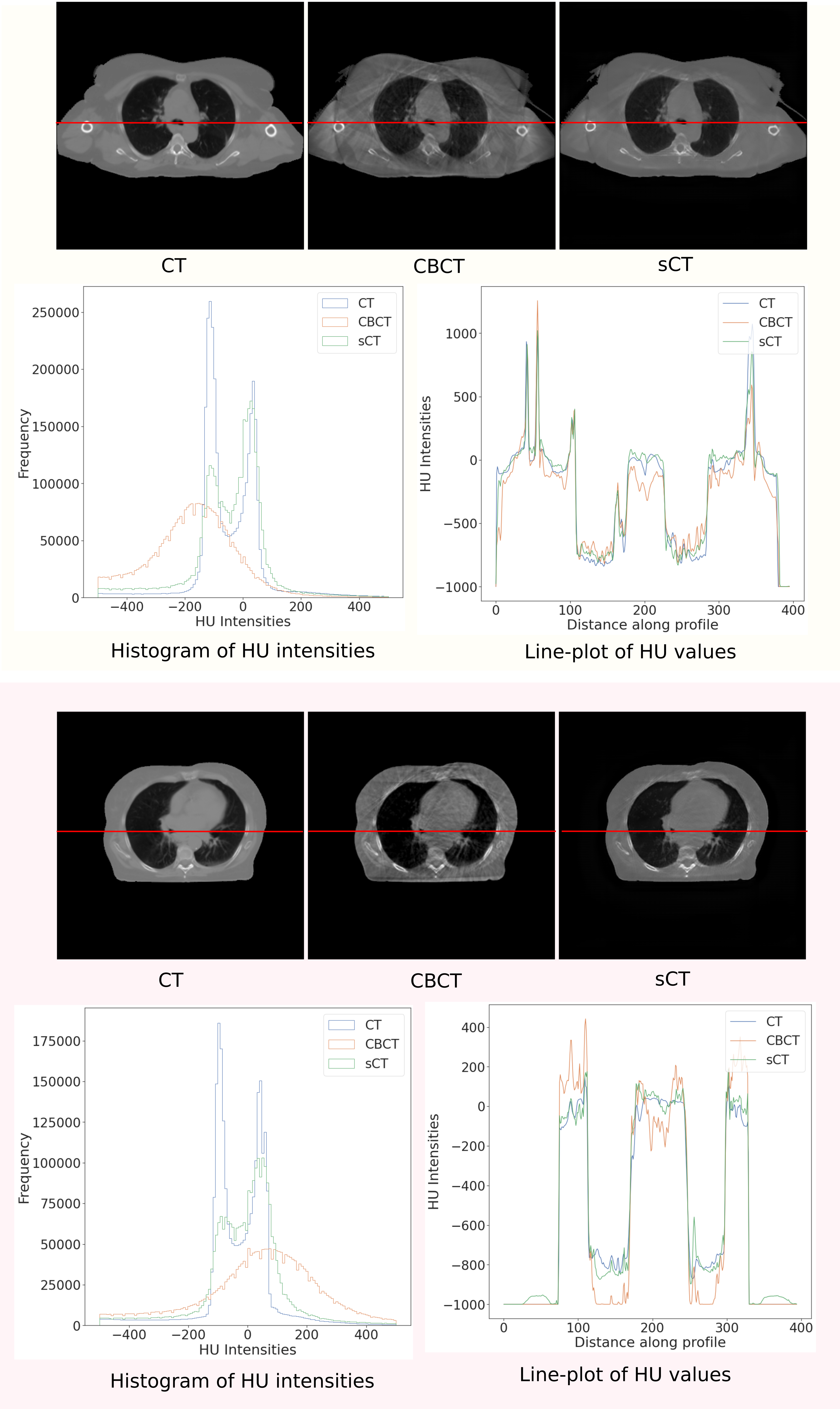}
    \caption{Histogram of HU intensities and line-profiles shown for CT, CBCT and sCT on two patients chosen randomly from the test set. The line chosen for profiling is highlighted by a red line passing the axial view of the images.}
    \label{fig:results:histogram}
\end{figure}

Figure \ref{fig:results:histogram} shows the histogram of HU intensity values between $(-500, 500)$ on the full scan for CT, CBCT and sCT. In addition, it also shows the line profiles for the same set of scans. The line chosen for the profile is drawn in red over the scans.
For both the patients, the sCT calibrates well with the CT in  in terms of distribution of HU values in the soft-tissue region, made easily observable through the windowing. The sCT also matches the CT line profiles better compared to the CBCT. This behaviour extends across all the patients in the test set. 

\subsection{Automated Segmentation Task}
\begin{figure}
    \centering
    \includegraphics[width=\textwidth]{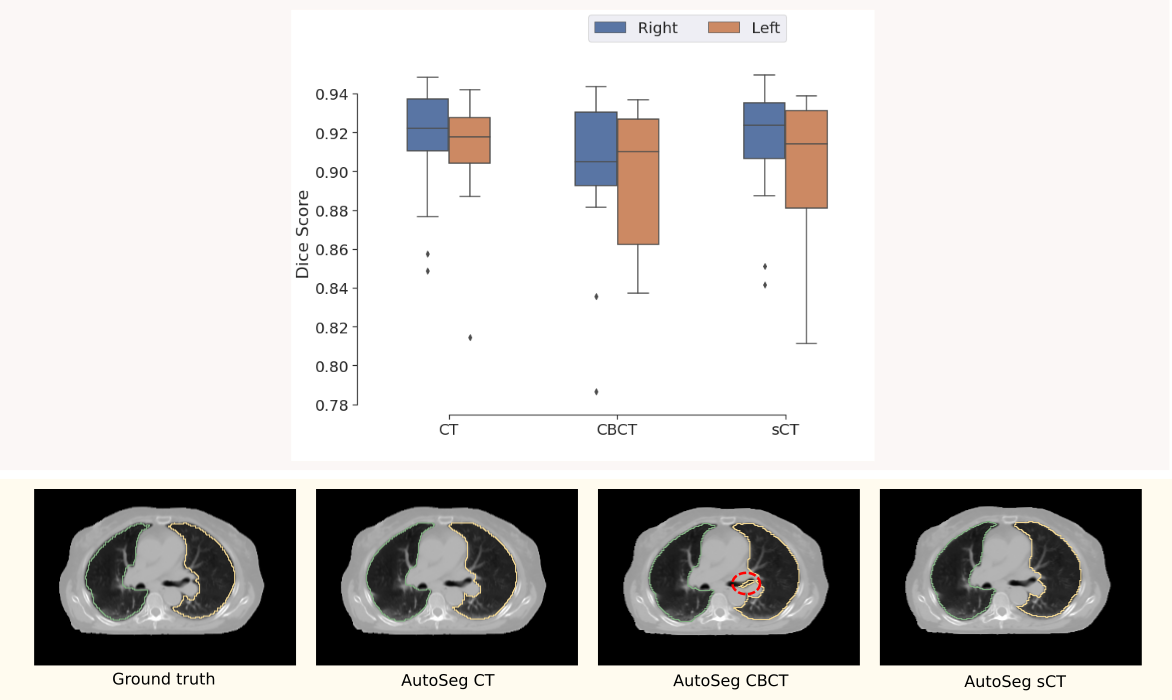}
    \caption{First row shows box plot showing Dice scores for segmentation of the right and left lung using CT, CBCT and sCT images as input. Second row shows visuals of segmentations generated by using different inputs to the automated segmentation pipeline, namely CT, CBCT and sCT. The ground truth segmentation, contoured by a clinician, is also shown. Note that all segmentations are shown on the CT as the original segmentations were drawn on the CT.}
    \label{fig:results:seg_viz}
\end{figure}

The sCT is also evaluated on a downstream task of lung segmentation as described in \ref{sec:methods:domain_eval}. The CT, CBCT and sCT are contoured for left and right lung and compared with ground truth contours for the same, available in the dataset. Figure \ref{fig:results:seg_viz} shows the box-plot of Dice scores obtained across all patients along with ground truth and automated contours generated on a randomly selected patient. Table \ref{tab:results:mean_dice} shows the mean Dice scores. 
\begin{table}[H]
\centering
\begin{tabular}{lll}
\hline
     & Left Lung         & Right Lung        \\\hline
CT   & \textbf{0.910984} & 0.913132         \\ 
CBCT & 0.898207          & 0.902951          \\
sCT  & 0.900434          & \textbf{0.914695} \\
\hline
\end{tabular}
\caption{Mean Dice scores for left and right lung segmentations for the CT, CBCT and sCT images as input to the automated segmentation model.}
\label{tab:results:mean_dice}
\end{table}
From Table \ref{tab:results:mean_dice}, we see that the sCT segmentations, on average, provide improved Dice scores when compared to CBCT segmentations. The mean Dice on the sCT, even improves over the CT for the right lung. The visualization of segmentation contours in Figure \ref{fig:results:seg_viz} shows a sample case where the CBCT is worse than the sCT (highlighted in red). Note that similar behaviour, where CBCT misses/adds parts of the contour, is seen in contours generated across multiple patients in the test set. 

\chapter{Discussion}
In this section, we discuss the results obtained from various experiments conducted and attempt to connect them to provide answers to the research questions posed in Section \ref{sec:intro:contrib}. 

Section \ref{sec:expts:RQ1} presents results for experimenting with different values of existing CycleGAN parameters and aims to answer the first research question of sufficiency of existing constraints. For \textit{map $\leftrightarrow$ aerial photo}, we identified that the default parameters provide the best results. However, the results are underwhelming as they fail to retain complex structure present in the aerial imagery. It seems that unpaired methods, in their existing state, may have trouble capturing such structural complexities. This is also corroborated through results from the original CycleGAN paper \citep{Zhu2017}. On the use-case of CBCT to CT, the structural complexity is vastly reduced in comparison to aerial imagery, and much more acceptable results are obtained using the default CycleGAN parameters. However, we identify failure modes such as loss of quality through addition of checkerboard artifacts and remnants of scattering artefacts in the generated images. Although it remains to be evaluated as suitable for clinical use, it does not offer the CT-like properties, especially in terms of smoothness. 

Experiments and results associated with adaptation of constraints were shown in Section \ref{sec:expts:RQ2}. For \textit{map $\leftrightarrow$ aerial photo} dataset, clear benefits were not observed by adapting constraints, while it was even detrimental for the CBCT-CT dataset. This is likely due to difficulty in associating control signals to a strategy that effectively adapts constraints. As described in Section \ref{sec:methods:adaptive_control}, multiple strategies were explored to go from sensing control signals to using that information to adapt $\lambda$ values. This was extremely challenging due to the large space of possibilities on when and how to adapt the parameters. This association is even more challenging for 3D models as they are trained for lesser number of iterations which might lead to design of more aggressive adapting strategies. The discriminator-based control signal completely failed in driving $\lambda$ changes while the cycle-consistency control signal was better, possibly due to the direct relationship between the cycle-consistency control signal and the constraints. 
As experimentally validated with the discriminator-based control signal, a single control signal may not always directly advise a change in constraints, therefore, better ways to indirectly use them to adapt parameters need to be investigated. To address this, looking at more than one control signal at a time was a potential idea. For instance, looking at the cycle-consistency signal along with variance of adversarial losses as a second control signal. However, it was dropped due to the large number of experiment configurations that would need to be investigated. More appropriately, multi-objective optimization methods such as hyper-volume optimization may prove to work well for these methods since they deal with balancing constraints in a more optimal way than simple additive combination. 

Various data-driven constraints and their results on translation are shown in Section \ref{sec:expts:RQ3}. We demonstrated that losses built for other medical imaging tasks may not work too well when introduced into the CycleGAN framework. Although MIND performed satisfactorily on quantitative scores, it rendered the images unusable due to  large modifications made to patient anatomy. Registration losses, that seem to offer properties that would benefit structure retention, as presented in Section \ref{sec:methods:reg_loss}, failed to converge across multiple settings when combined with the rest of the constraints in the CycleGAN. In contrast, simple frequency-based losses seem to combine well with existing constraints in the CycleGAN and provide translations with desirable qualitative and quantitative scores. Combination of MIND with frequency-based loss seemed to get rid of the GAN-induced artifacts but the combination performed poorly on quantitative scores and showed other qualitative issues. 

Another major observation is the insufficiency of solely relying on quantitative analysis in choosing the best model. This is observed even with out-of-distribution data where strong pairs were formed. For instance, Frequency $L1$ provides one of the worst scores on the phantom but is superior to the other models as it does not induce artifacts or result in reduction in quality that all other models were susceptible to. This puts forward the question if existing image similarity metrics can be relied on fully to evaluate such methods. Research into evaluation methods that can sufficiently capture these properties in generated images would push the field closer toward general and clinical acceptance. \cite{Gragnaniello2021} present a review of existing methods for synthetic image detection and propose potential research areas for the future. These methods could also help in quantitatively determining undesirable additions such as artifacts in the generated images. 

Domain-specific methods of evaluation can provide good insight into the clinical usability of a particular set of methods. As seen in Section \ref{sec:expts:RQ4}, synthetic CT generated from the best-performing model provided HU intensity distribution and line profiles inline with real CTs. Automated segmentation on the synthetic CT showed performance on par with real CT (even better for the right lung) and improvements from the original CBCT. Given the simplicity of the translation process, it can be integrated into existing clinical workflows to improve the quality of the CBCT. The improved CBCT can be useful for multiple downstream tasks, from improving auto-contouring to adapting treatment plans. 
\chapter{Conclusion}

In our work, we investigated various constraints that affect CycleGAN-based CBCT enhancement along with supporting methodology and evaluation. We began by \textit{exploring existing CycleGAN constraints to determine if they are sufficient for generating high-quality images, in order to answer our first research question.} Through our qualitative and quantitative evaluations, we conclude that although an improvement in calibration is observed when compared to the original, the image quality is impaired through the addition of minor artifacts and decreased resolution. As this is undesirable for clinical acceptability, we looked at possible solutions through answering the subsequent questions that the thesis proposed. \textit{In order to answer our second research question, we designed adaptive control strategies to dynamically balance cycle-consistency and adversarial losses,to gauge their impact on generating improved CBCT images.} However, this approach was not successful in improving image quality, demonstrated both qualitatively and quantitatively. We then reported the complexity of designing such strategies, due to the large space of control signals and adaptive strategy choices that need to be made, and subsequently proposed more optimal multi-objective methods as future work. Finally, we \textit{looked at data-driven constraints, in the form of structure-consistency losses, as part of investigating our third research question}. This investigation proved to be most successful, with the best method, the generalized frequency loss, improving MAE, MSE, NMSE, and PSNR, by 4\%, 20\%, 18\%, and  3\%, respectively, compared to the baseline. The generalized frequency loss, implemented as a part of this thesis, proved to not only improve over the baseline, but also outperform existing methods, such as the MIND loss \citep{yang2018unpaired}. This was done at a much lesser memory and compute cost.  More importantly, in terms of qualitative comparison, we observed the best performance, with no drops in image quality or any addition of artifacts. \textit{As a part of answering our fourth research question, all our evaluations were conducted using image-similarity based quantitative evaluation and structured qualitative evaluation.} We leveraged weak pairs of patient data and a strong pair of phantom data obtained through image-registration. The phantom data was also used as out-of-distribution data. Efficacy of our methods were corroborated by subjecting them to out-of-distribution data, where they demonstrated sufficient accuracy and did not provide unexpected results. \par
One of the core goals of this thesis was providing a more reliable and robust method for CBCT improvement. This improved CBCT can benefit various adaptive radiotherapy workflows in the clinic such as auto-contouring, image registration, and dosimetry. Improvements in these worflows can save clinicians' valuable time and effort along with reduction of costs associated with repeated imagery. Improved CBCT can not only benefit the clinic, but also the patient as improved quality CBCTs can mean lesser CT scans and therefore, lesser radiation exposure. \textit{In an extension to the fourth research question, we implemented clinically motivated evaluations such as HU intensity distribution comparisons and line profiles}, where we demonstrated that the improved CBCT matches the fan-beam CT very well. Furthermore, the value of improved CBCTs in down-stream tasks was shown through a comparison of contours generated through lung auto-segmentation. Mean Dice scores of contours on the improved CBCT were comparable to the fan-beam CT and surpassed the CBCT. We point out that a full dosimetric evaluation would help better establish clinical applicability of our methods. However, due to complexity of designing clinically acceptable treatment plans, we exclude this in the thesis and propose it as future work. Another promising line of future work is the development of quantitative methods that can capture artifacts in generated images so as to decrease the reliance on subjective qualitative inspection, which is heavily utilized in this thesis.

\begin{appendices}
\begin{figure}[H]
    \centering
    \includegraphics[width=\textwidth]{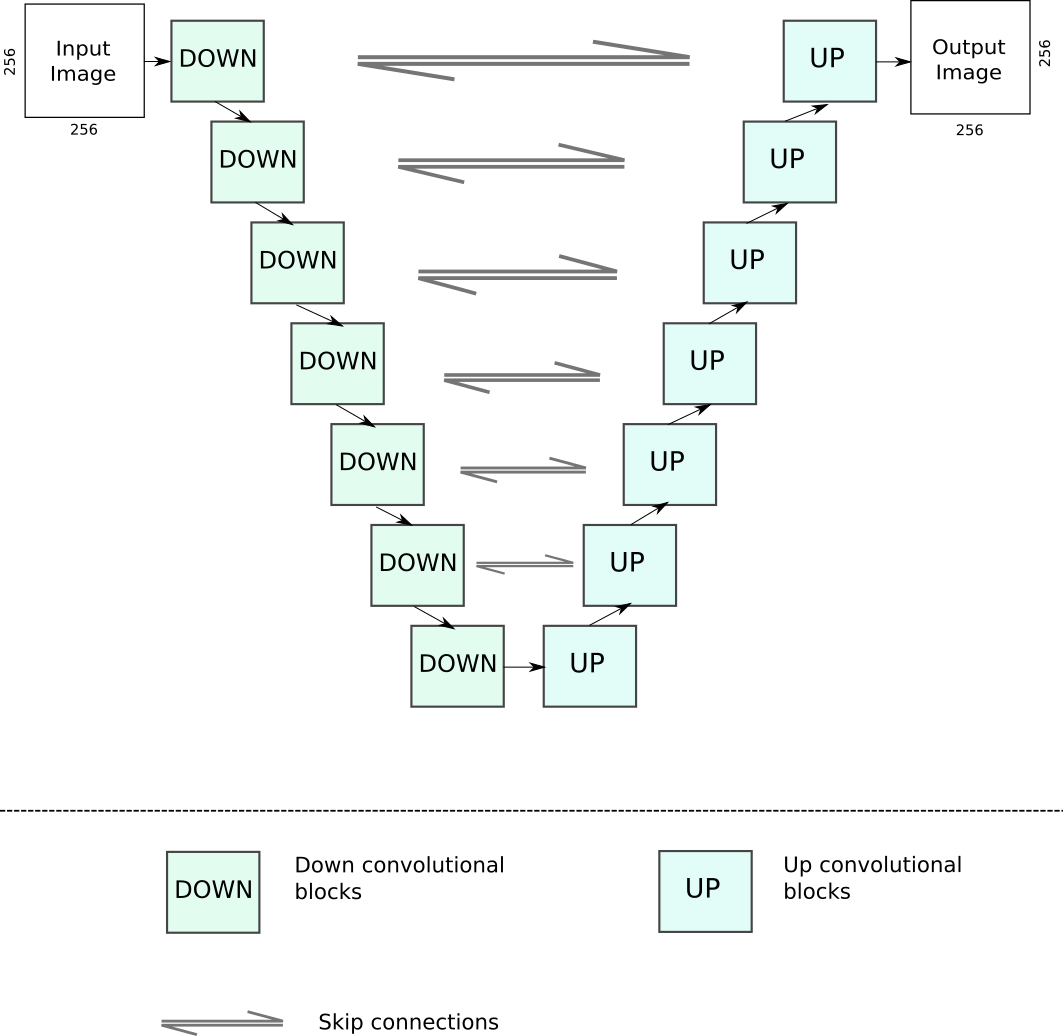}
    \caption{Block diagram of the 2D base configuration architecture. It consists of 7 down and up convolutional blocks with skip connections as described in Section \ref{sec:expts:2d_conf}.}
    \label{fig:appendix:2d}
\end{figure}

\begin{figure}[H]
    \centering
    \includegraphics[width=\textwidth]{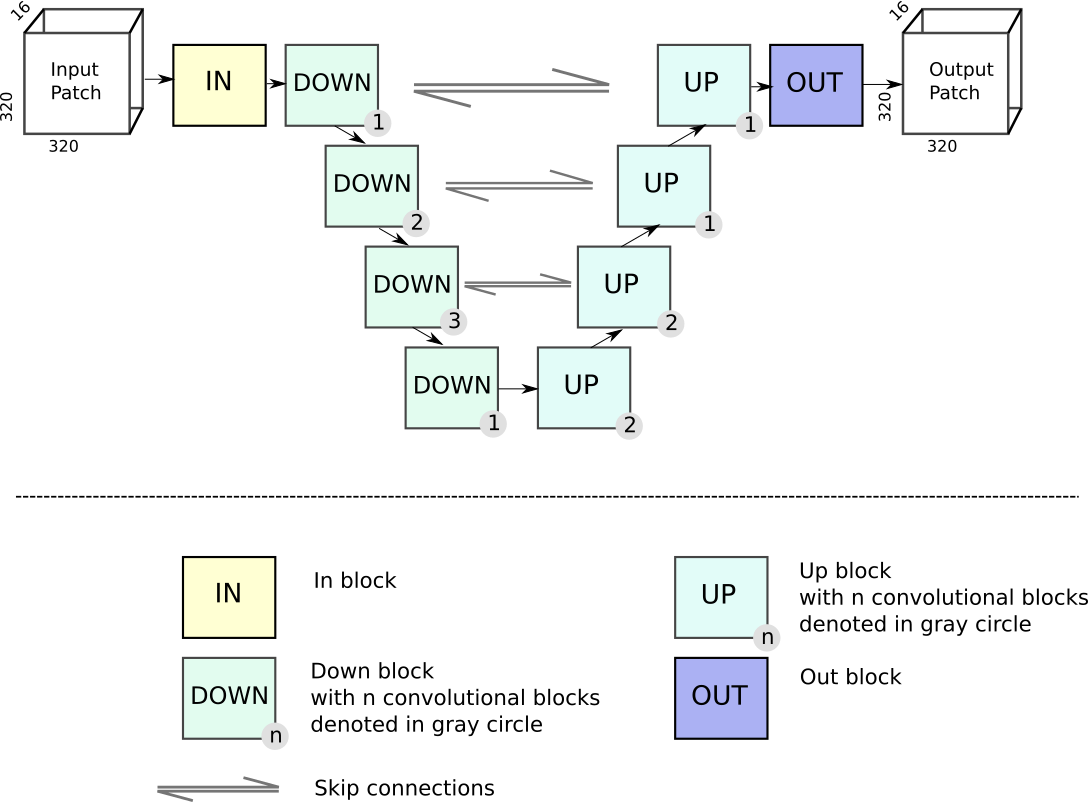}
    \caption{Block diagram of the 3D base configuration architecture. It comprises of input blocks, output blocks, up blocks and down blocks along with skip connections as described in Section \ref{sec:expts:3d_conf}.}
    \label{fig:appendix:3d}
\end{figure}

\end{appendices}

\end{document}